\documentclass[aip,sd,amsmath,amssymb,reprint]{revtex4-1}
\usepackage[utf8x]{inputenc}
\usepackage{textcomp}
\usepackage{graphicx}
\usepackage{epstopdf}
\usepackage{bm}
\usepackage{caption}
\usepackage{subcaption}
\usepackage{multirow}
\usepackage{color,soul}

\usepackage[colorinlistoftodos,prependcaption,textsize=tiny]{todonotes}

\begin{document}


\title{Nonlinearity in stock networks}

\author{David Hartman}
\email{hartman@cs.cas.cz}
\affiliation{Institute of Computer Science, Czech Academy of Sciences, Prague, Czech Republic}

\author{Jaroslav Hlinka}
\email{hlinka@cs.cas.cz}
\affiliation{Institute of Computer Science, Czech Academy of Sciences, Prague, Czech Republic}

%

\date{\today}

\begin{abstract}

Stock networks, constructed from stock price time series, are a well-established tool for the characterization of complex behavior in stock markets. Following Mantegna's seminal paper, the linear Pearson's correlation coefficient between pairs of stocks has been the usual way to determine network edges. Recently, possible effects of nonlinearity on the graph-theoretical properties of such networks have been demonstrated when using nonlinear measures such as mutual information instead of linear correlation. 
In this paper, we quantitatively characterize the nonlinearity in stock time series and the effect it has on stock network properties. This is achieved by a systematic multi-step approach that allows us to quantify the nonlinearity of coupling;  correct its effects wherever it is caused by simple univariate non-Gaussianity; potentially localize in space and time any remaining strong sources of this nonlinearity; and, finally, study the effect nonlinearity has on global network properties. By applying this multi-step approach to stocks included in three prominent indices (NYSE100, FTSE100 and SP500), we establish that the apparent nonlinearity that has been observed is largely due to univariate non-Gaussianity. Furthermore, strong nonstationarity in a few specific stocks may play a role.  In particular, the sharp decrease in some stocks during the global financial crisis of 2008 gives rise to apparent nonlinear dependencies among stocks.

\end{abstract}

\pacs{Valid PACS appear here}
\keywords{stock network, nonlinearity, mutual information, complex network, graph theory}
\maketitle

\begin{quotation}
Stock networks have received increasing attention in the scientific literature over the last two decades. Motivated mainly by portfolio optimization, risk management, financial policy making, and crisis and extreme events characterization, stock networks have been analyzed from various perspectives, alongside different methods of filtering or preprocessing of stock time series. Still, the majority of studies assume only linear relations to be present in the dependencies between stocks, and accordingly use Pearson's correlation coefficient to determine the weights of network edges. Only recently have some authors applied nonlinear measures and have established different stock network properties, supposedly due to nonlinearity being undetected by linear correlation. We demonstrate in this study that the effects that have been described by these authors is likely to be not due to nonlinearity but largely to marginal non-Gaussianity and nonstationarity in the stock time series. We suggest a method to minimize these effects via a specific preprocessing pipeline that would mitigate these components of supposed nonlinearity and localize the remaining effects of different, possibly true nonlinear effects. We believe that this study can significantly help researchers in the field of stock network analysis by providing an explanation of various effects that emerge when constructing stock networks. Moreover, our approach might be applicable in other disciplines utilizing networks constructed from time series.
\end{quotation}

\section{Introduction}
Many real-world systems, such as the human brain, the Earth's climate or metabolic subsystems, have a characteristic intrinsic connectivity structure among their subsystems. This structure enables specific characterization of a corresponding system using connectivity information represented by what are known as complex networks~\cite{Boccaletti2006,BullmoreSporns2009complex,Jeong2000}. As well as such systems appearing in nature, there are examples involving human activities, such as general social networks~\cite{zachary1977information}, the Internet~\cite{Albert1999}, and email networks~\cite{guimera2003self}. 

The characterization of systems using complex networks has many applications across diverse fields, ranging from clinical purposes in brain studies~\cite{He2008structural,Liu2008}, through disentangling various climate system states~\cite{Wiedermann2016climate}, analyzing collective behavior of social groups~\cite{WassermanFaust1994social}, understanding protein interactions~\cite{Jeong2001lethality}, to the analysis of resilience of the Internet to random breakdowns~\cite{Cohen2000resilience}. The reviews of complex network applications~\cite{Costa2011analyzing,Havlin2012challanges} contain further examples. 

In the financial domain, which is of interest in this work, there are two systems commonly studied using complex networks: world-trade networks~\cite{SerranoBoguna2003topology} and stock markets~\cite{Mategna1999hierarchical}. In this work we deal specifically with stock networks. There are various applications of stock networks, such as the design of a trading strategy and its corresponding portfolio optimization via the network characterization of the Markowitz portfolio~\cite{Onnela2003dynamics,Tola2008cluster}; localization of new investment based on topological properties of the corresponding network~\cite{Peralta2016network}; using specific topological properties of networks to evaluate systematic risk~\cite{Meng2014systematic}; alternative validation of market models~\cite{Bonannno2004networks}; and characterization of extreme events such as a financial crisis~\cite{Heiberger2014, Heiberger2018}.

Most applications of complex networks use their topological properties. A commonly discussed example is the global property of complex networks called small-world~\cite{Watts1998}. Despite some criticism of its definition~\cite{Hlinka2012smallworld,Papo2016,Hlinka2017smallworld}, numerous studies have used the small-world property for the characterization of phenomena in real-world systems. Examples include atmospheric teleconnection patterns due to their long-range connections~\cite{tsonis2006networks,phillips2015graph} in climate science; emergence of a small-world structure in the human brain system analyzed via functional magnetic resonance imaging (fMRI) data~\cite{Heuvel2008a}; and stock networks having small-world effects at different time horizons~\cite{Tumminello2007correlation}. As well as global properties such as small-world, local properties that relate to particular network nodes are also commonly used, for example for the purposes of prediction~\cite{Caraiani2017predictive}.

Regardless of the character of the network property being studied, a first crucial step in network analysis is construction of the network. In this step, a network structure should be defined in a form that reasonably represents the interdependence among existing subsystems of the studied system. 
Some systems have their link weights defined naturally from the data, as in the case of world-trade networks~\cite{SerranoBoguna2003topology} where nodes are different countries in the world and weights are given by the amount of trade between these countries, or structural human brain networks where links between cortical areas are given by the physical presence of white matter tracts between them. This type of connectivity is sometimes called {\em structural connectivity} in contrast to {\em functional connectivity}, which in the case of the brain network is defined as the statistical dependencies between remote neurophysiological events~\cite{Friston1994}. A similar principle of defining network links based on the dependency between observed time series is typically used in stock networks.

Functional connectivity is thus generally given as the dependence between time series or appropriately defined sub-series, such as by a sliding time window~\cite{Onnela2003dynamics}. One of the common measures used within stock networks is Pearson's correlation coefficient~\cite{Tumminello2007correlation} computed for the logarithmic return of time series of adjusted closing prices after any necessary preprocessing. 
Pearson's correlation coefficient is also commonly used in other areas, such as brain networks~\cite{BullmoreSporns2009complex} and climate networks~\cite{tsonis2004architecture}. For brain networks there has been increasing interest in the potential of using measures accounting for nonlinear dependence, such as mutual information~\cite{Stam2005Nonlinear}. It has been shown, however, that with appropriate preprocessing the data from common brain imaging modalities, such as functional magnetic resonance imaging, may be, in fact, very close to Gaussian~\cite{Hlinka2011NeuroImage} and the construction of the complex network in this case and its subsequent analysis is not biased when using linear measures~\cite{Hartman2011Chaos}. For climate complex networks, most of the supposed nonlinearity can be also removed via proper preprocessing and any remaining pieces can be readily located~\cite{Hlinka2013ClimDyn}. Note that all the above-mentioned networks are undirected, where the characterization of possible nonlinearity can be based on such measures as mutual information. For directed networks, a proper characterization of the contribution of nonlinearity is a more challenging task; see examples of handling nonlinear dependence in directed climate networks~\cite{Hlinka2013Entropy}.

Recently, for stock networks there have been some preliminary results concerning the use of nonlinear measures based on mutual information and related principles~\cite{Fiedor2014networks}. The author argues that using nonlinear measures, in particular mutual information and mutual information rate, may be suitable for network analysis of stock data, due to the potential nonlinearity of the dependencies.
To support his claims, the author analyses the top 100 U.S. stocks traded on the New York Stock Exchange (NYSE),  included in the NYSE100 index maintained by the New York Stock Exchange~\cite{nyse}, and reports differences in graph-theoretical characteristics -- for example, average path length and betweenness centrality -- compared to networks constructed using a linear correlation coefficient. 

This topic of nonlinearity in network construction has attracted more researchers. Rocchi et al. have explored properties of complex networks using a version of transfer entropy~\cite{Rocchi2017emerging} when considering a set of 100 stocks traded on the London Stock Exchange with the highest capitalization, i.e. those stocks featuring in the Financial Times Stock Exchange 100 (FTSE100) index, maintained by the FTSE Group~\cite{ftse100}. 
Using nonlinear measures has recently also played a role within one of the dominant areas of application, portfolio optimization. One of the first studies of portfolio optimization and the design of investment strategies using a complex network constructed with weights computed from mutual information is the work of Kaya~\cite{kaya2015eccentricity}. In this study, mutual information is used to compute connectivity weights and to construct an unweighted network using Minimal Spanning Tree (MST) filtering. Subsequently, for each node of this graph the author computes the eccentricity, i.e. the maximal distance to any other node. This is used to characterize extreme events, and to create a tool for portfolio optimization. The author does not compare linear and nonlinear approaches, only mentioning various advantages and drawbacks of either approach in the introduction to his paper. 

Quite recently, Baitinger and Papenbrock compared Pearson's correlation coefficient and mutual information for the purposes of portfolio optimization, expanding on their previous work~\cite{Baitinger2017interconnectedness} that used only linear measures. They construct MST networks for $28$ chosen assets for the time period $2000 --2015$ and compare correlation and mutual information using various approaches. Their dataset is not directly comparable to using only stocks from a particular index, as their ``multi-asset dataset'' contains not only stocks but also commodities, currencies, etc. In their initial observations they already mention that, by visual inspection, correlation and mutual information values seem to be close even to the extent that a quadratic or polynomial fit could explain it. 
In second part of their work the authors study the similarity of MST graphs constructed using either correlation or mutual information. As a similarity measure they use a variant of the Jaccard index, i.e. the number of pairs of nodes with different edge/no-edge presence, and what is known as PCA centrality, i.e. the first principal component of PCA calculated for a set of centralities~\cite{Baitinger2017interconnectedness2}. They observe that despite the above-mentioned similarities, the networks are only moderately related. 

Despite the interest and potential importance of the question of the contribution of nonlinear dependence in stock networks, results reported so far have been generally limited to statements concerning the close connection between both dependence measures~\cite{Baitinger2017interconnectedness2}, their theoretical differences~\cite{kaya2015eccentricity,Fiedor2014networks}, and the numerical observation of different properties of networks constructed from linear correlation and mutual information~\cite{Baitinger2017interconnectedness2,Fiedor2014networks}. However, general statistical reasoning as well as practical experience from other applications of network science~\cite{Hlinka2011NeuroImage,Hartman2011Chaos, Hlinka2013ClimDyn} suggest that divergence between mutual information and linear correlation networks might be to a large extent due to random variation arising from the estimation of quantities from finite size samples, as well as due to nonstationarities in the underlying process. This establishes the need for a deeper assessment of the effect of nonlinearity in stock networks, including its strength, localization, origin and effect on network properties. To the best of our knowledge, nonlinearity has not been tested to this level of detail in the context of stock networks. 

The paper has the following structure: Section~\ref{s:data} contains a description of the data; Section~\ref{s:methods} describes the methods used in our analysis, including preprocessing, connectivity determination, and the network measures considered; Section~\ref{s:nonlinass} contains a description of the suggested pipeline, providing in parallel the corresponding results of the analysis for the NYSE100 index to enhance readability of the paper; and Section~\ref{s:discussion} contains a discussion and concluding remarks. Results of further analysis, including extension to other stock indices (FTSE100 and SP500), are included in the Supplementary material.

\section{Data}\label{s:data}

We use historical stock prices downloaded from Yahoo! Finance service~\cite{YahooFinance2017}. In particular, we include stocks belonging to the New York Stock Exchange 100 (NYSE100) index, maintained by the New York Stock Exchange~\cite{nyse}; see list of stocks in Table~1 in the Supplementary material. We consider only stocks traded between 11 November 2003 and 7 November 2013. This restriction results in $N = 89$ stocks, as can be observed in Figure~1 in the Supplementary material to this text. For daily data this leads to a data length of $T = 2608$.
We also replicate the key analysis on stock indices included in the Financial Times Stock Exchange 100 index (FTSE100), maintained by the FTSE Group~\cite{ftse100}, and the set of 500 American stocks with highest market capitalization traded at NYSE or NASDAQ, included in the Standard \& Poor 500 (SP500) index, operated by S\&P Dow Jones Indices~\cite{sp}.

From the initial data we used the daily adjusted closing price that accounts for any possible divides and splits of stocks. These data are processed in a standard way to account for missing values using linear interpolation; the maximum number of missing days was $2$ and missing data were relatively rare.

The last processing step is the computation of the {\em logarithmic return}, see for example~\cite{Mategna1999hierarchical}. Let $p_i(t)$ be the (adjusted) closing price of stock $i$ at time $t$ (in our case, time is given in days). Then the logarithmic return of stocks $r_i(t)$ is defined as

\begin{equation}
r_i(t) = \log\left[ \frac{p_i(t)}{p_i(t-1)}\right].
\end{equation}

Note that the transformation above corresponds to taking first differences of the log-transformed prices, i.e. a relatively simple data transformation common in general time-series analysis. Following the standard approach, this characteristic is used for network construction~\cite{Mategna1999hierarchical}.

\section{Methods}\label{s:methods}

We present a collection of methods to handle nonlinearity alongside a standard process of stock network analysis. Settings of dates and parameters of estimators are inspired by recent work concerning nonlinearity in stock networks~\cite{Fiedor2014networks}. There are basically two important steps within most complex network analyses, which this section describes along with the impact of nonlinearity and its handling. First, graph connectivity determining weights of edges has to be computed from the data with appropriate preprocessing. Secondly, a network is constructed and various graph characteristics are computed. These processes are covered in the following two sections.

\subsection{Graph connectivity}

The first step in unweighted network construction is the determination of edges between nodes. This task is performed by computing the weights of edges and their subsequent filtering. The weight of an edge between two stocks is commonly estimated as the correlation between the logarithmic returns of their adjusted closing prices using Pearson's correlation coefficient. For two real random variables $X$, $Y$, this coefficient is defined by

\begin{equation}
\rho_{X,Y} = \sqrt{\frac{E[(X - E(X))(Y - E(Y))]}{E[(X - E(X))^2] E[(Y - E(Y))^2]}},
\end{equation}

where $E(\cdot)$ denotes expected value. We denote finite estimates of the correlation coefficient by $r(\cdot, \cdot)$. It is often useful to consider distances rather than weights. For this reason, and in accordance with the approach in~\cite{Fiedor2014networks}, we transform the computed correlations using the Euclidean metric~\cite{Mantegna1997degree}:

\begin{equation}
\delta_{X,Y} = \sqrt{2(1-\rho_{X,Y})}.
\end{equation}

We use mutual information to test any nonlinearity in the interdependence. For two discrete random variables $X$ and $Y$ with values drawn from $\mathcal{X}$ and $\mathcal{Y}$, the mutual information is defined by

\begin{equation}
I(X,Y) =  \sum_{y\in \mathcal{X}}\sum_{x\in \mathcal{X}} p(x,y) \log \frac{p(x,y)}{p(x)p(y)},
\end{equation}

where $p(x) = Pr[X = x], x\in \mathcal{X}$ is the probability distribution of random variable $X$ and $P(x,y) = Pr[(X,Y) = (x,y)]$ is the joint probability distribution of $X$ and $Y$. For continuous random variables, we define mutual information with using integration instead of summation and estimate its value using an appropriate discretization. 

There are various methods to estimate mutual information depending on the way a continuous probability space is discretized. A common choice for these estimates is a simple box-counting algorithm based on the marginal equiquantization method~\cite{Palus1993information}, which generates partitions such that marginal bins are equiprobable. For the time series length under consideration, 4 bins have been previously suggested for discretization~\cite{Fiedor2014networks,Fiedor2014Frequency} is $4$, although this author also mentions the possibility of using $8$ bins, based on another reference~\cite{NavetChen2008predictability}. This choice is also below the established bound provided in the literature~\cite{Palus1995testing}, where the author suggests using a number of bins strictly less than $Q = \sqrt[n+1]{T}$, which for $n_{var} = 2$ variables and length of time series $T = 2608$ equals to $13.765$, making both choices of $4$ and $8$ acceptable. 

Estimation of mutual information from a finite sample suffers from bias dependent on sample size. The correction of this bias has been previously suggested~\cite{Hlinka2011NeuroImage}, and has later been used to correct input data for complex networks~\cite{Hartman2011Chaos,Hlinka2013ClimDyn}. The idea of this correction is the comparison of estimated mutual information values with a population of realizations having the same sample size and analytically established values of mutual information. For this reason, the first step in our analysis is altering the computation of mutual information in order to account for this bias. We use this step for each estimation of mutual information. 



When the variables $X$ and $Y$ have a bivariate Gaussian distribution, we can compute the mutual information $I(X,Y)$ based on linear correlation using the following well-known equation:

\begin{equation}\label{eq:gaussmi}
I(X,Y) = I_G(X,Y) \equiv -\frac{1}{2}\log(1 - \rho^2_{X,Y}).
\end{equation}

For a general distribution which is not Gaussian this equation may not hold. In the following, we use the term nonlinearity instead of non-Gaussianity to stay in line with the literature. Importantly, nonlinearity can be caused by two reasons. Firstly, for bivariate dependence nonlinearity can be given by non-Gaussianity of univariate marginal distributions, which does not affect mutual information, but may change correlation values substantially. Secondly, even when there is a normal marginal distribution, nonlinearity can be caused by non-Gaussianity of the bivariate dependence pattern, i.e. the copula.

In cases where nonlinearity is not only hidden in the marginal distributions, we can evaluate the non-Gaussian information using a recently suggested method~\cite{Hlinka2011NeuroImage}. This method is based on the simple fact that variables $X$ and $Y$ with univariate Gaussian distribution have their mutual information bounded from below by the Gaussian mutual information computed from linear correlation using Equation~\eqref{eq:gaussmi}~\cite{Hlinka2013ClimDyn}. More formally, this is to say that $I(X,Y) \geq I_G(X,Y)$, in which equality is acheived when the bivariate dependence is Gaussian. In this way, we can define the {\em extra-normal} ({\em non-Gaussian}) information~\cite{Hlinka2013ClimDyn} as the deviation of mutual information in the original data from the mutual information in a corresponding bivariate Gaussian distribution:

\begin{equation}\label{eq:extranorm}
I_E(X,Y) \equiv I(X,Y) - I_G(X,Y)
\end{equation}

This characteristic can be used to quantify the amount of nonlinearity in the data.



\subsection{Network properties}

A {\em network} is an unweighted graph $G = (V,E)$, where $V$ is a set of nodes and $E$ is a set of (unweighted) edges. Similarly to the number of stocks defined above, the size of a graph is defined as the number of nodes and denoted by $|V(G)| = N$. Occasionally it is necessary to define a weighted graph, which we understand as a triple $G_w = (V,E,w)$, where $V$ and $E$ are defined as before and $w:E \rightarrow \mathbb{R}$ is the weight function assigning a weight to each edge. Note that the original matrix of correlation or mutual information values constructed in the first step from the data is symmetric and represents a collection of weighted edges and thus a weighted graph. We define the density of a graph as the number of edges relative to the maximum possible number of edges, i.e. $\rho_G = 2|E| / N(N-1)$. It is sometimes important to be able to describe characteristics of specific parts of a graph. For vertices and edges we use standard set notation. For a set of nodes $V' \subset V(G)$ we write $E(V')$ for all edges $E' \subset E(G)$ such that for $e = \{u,v\} \in E'$ it holds that $u,v \in V'$. The set of edges $E(V')$ define a graph $G' = (V',E')$, equal to the subgraph of graph $G$ induced by the set $V'$. A commonly used subset of vertices here is the neighborhood of a node $v$, denoted by $\Gamma(v)$, which consists of all nodes $u \in \Gamma(v)$ such that $\{u,v\}\in E(G)$.

Since we understand a network as an unweighted graph, we need to binarize the matrix of weights to obtain an unweighted graph from the original weighted one. There are various standard strategies. One of the oldest and most commonly used strategy for stock networks is to use the Minimal Spanning Tree (MST) of the original weighted graph~\cite{Mategna1999hierarchical}, for which we use the standard Kruskal algorithm, leading to a simple network, i.e. a tree (a connected graph with no cycles). Reduction of edges from the original weighted graph is quite drastic by this method~\cite{Tumminello2005tool}. That is why another approach has been suggested, called Planar Maximally Filtered Graph (PMFG)~\cite{Tumminello2005tool}. This approach finds an unweighted graph which keeps as many high-weighted edges as possible while remaining planar. Our algorithm starts with an empty copy of the original graph, i.e. without edges. Further we sort the original edges reversibly according to their weights and iterate them in this order, i.e. from the maximum weight to the lowest one. At each step, we try to add an edge to the currently constructed graph and decide whether the resulting graph is planar. If not, the edge is removed. Both these approaches, i.e. MST and PMFG, are without any parameters and result in connected graphs that, in some sense, keep only the necessary strongest edges.  The winner-take-all (WTA)~\cite{Tse2010network} approach is parametric and based on keeping edges according to a given threshold for weights, i.e. keeping all edges having weight greater than or equal to a given threshold. One of the disadvantages of this approach is the possibility of producing disconnected graphs. 

For an undirected graph, we can define several characteristics that can be further used to analyze the underlying system, see for example~\cite{Boccaletti2006}. We have chosen a representative set of characteristics based on their popularity within stock networks literature, namely degree or degree centrality~\cite{Heiberger2018, Baitinger2017interconnectedness, Aste2010correlation}, clustering coefficient~\cite{Onnela2004, Boginski2005Statistical, Huang2009network}, closeness centrality~\cite{Heiberger2018, Baitinger2017interconnectedness, Aste2010correlation}, betweenness centrality~\cite{Heiberger2018, Baitinger2017interconnectedness, Aste2010correlation}, eigenvector centrality~\cite{Baitinger2017interconnectedness}, eccentricity~\cite{kaya2015eccentricity,Aste2010correlation,Baitinger2017interconnectedness}, characteristic path length~\cite{MingXia2015}, and assortative coefficient~\cite{Heiberger2014,MingXia2015}.

We can roughly distinguish two types of characteristic -- local and global. Global characteristics evaluate the whole structure of the graph, while local ones characterize the structure around each node. 
An example of a local characteristic is node degree. For a node $v$, we define the degree $\deg_G(v)$ as the number of nodes adjacent to $v$ in graph $G$. We also use the standard shorter notation $\deg_G(v) = k_v$. The set of all node degrees is an important network characteristic, determining some fundamental complex network properties, such as scale-free character~\cite{Barabasi1999emergence}. Representing vertices as numbers from $[n]=\{1,2,\dots, n\}$, we define the {\em adjacency matrix} $A$ as $\{a_{u,v}\}_{u,v \in [n]}$, where $a_{u,v} = 1$ if $\{u,v\} \in E(G)$ and $a_{u,v} = 0$ otherwise. The {\em degree} of a vertex $v$ can then be defined as $k_v = \sum_{u \in [n] \setminus v} a_{v,u}$.

The small-world property~\cite{Watts1998}, mentioned above, is an example of a global characteristic. Roughly speaking, small-world is characterized by short average distances between nodes along with their high clustering. To determine distances, the shortest paths between nodes are used. A path is a sequence of nodes where two consecutive ones are adjacent in the graph and no vertex is present twice in this sequence. The length of a path is the number of edges in this sequence (one less than the number of nodes). The distance $d(u,v)$ between two nodes $u$ and $v$ is the length of the shortest path between them. Average distances are evaluated using {\em characteristic path length}, defined by
\begin{equation}
L = \frac{1}{n(n-1)}\sum_{u,v \in V(G)} d(u,v).
\end{equation}

Sometimes, a local version of this characteristic is used, called {\em node shortest path}, or {\em average shortest path}~\cite{Fiedor2014networks}. This characteristic expresses for each node $v$ the average distance from $v$ to any other node in the graph:
\begin{equation}
L(u) = \frac{1}{n-1}\sum_{v \in V(G) \setminus \{u\}} d(u,v)
\end{equation}

Note that $L(u)$ can be used to compute $L$. Let us also note that characteristics determined from distances may strongly depend on the connectedness of the graph. If a graph is not connected or even has many connected components, many distances can be set to infinity. There are various methods how to handle this situation. In the case of graphs obtained from filtering using either MST or PMFG this represents no problem as these are always connected -- given that we may assume that the original interdependence matrix is dense enough. In case of the WTA filtering we use the standard approach that computes averages only from existing paths.

The second main property characterizing the small-world property is the clustering of nodes. Clustering is characterized by the average tendency of a node to have a dense neighborhood. The measure used is the {\em clustering coefficient}, defined for a vertex $v \in (G)$ by
\begin{equation}
C(v) = \frac{2|\Gamma(v)|}{k_v(k_v - 1)} = \frac{\sum_{u,w} a_{v,u}a_{u,w}a_{w,v}}{k_v(k_v - 1)}.
\end{equation}

A global version of this characteristic, called the {\em average clustering coefficient}, is obtained by averaging local clustering, and is defined formally by 
\begin{equation}
C = \frac{1}{n} \sum_{v \in V(G)}C(v).
\end{equation}

The notion of a distance is used in many other types of characteristic. One important class of characteristics is that of centralities. One example is the {\em betweenness centrality}, which, roughly speaking, assigns to each node the relative level to which this particular node plays a role as a mediator when connecting other nodes via shortest paths. Formally, for a node $v$ choose all pairs of distinct nodes $u$ and $w$ and find out how many shortest paths connecting $u$ and $w$ contain the node $v$. Averaging these values across all pairs of nodes gives the {\em betweenness centrality}, i.e.

\begin{equation}
C_b(v) = \sum_{u,w\in V(G), v \neq u \neq w} \frac{\sigma_{u,w}(v)}{\sigma_{u,w}},
\end{equation}

where $\sigma_{u,w}$ is the number of shortest paths between $u$ and $w$ and $\sigma_{u,w}(v)$ is number of shortest paths between $u$ and $w$ passing through node $v$.

A simpler version of centrality using shortest distances is the {\em closeness centrality}. This characteristic calculates for each node the reciprocal value of the average distance to any other node, i.e.
\begin{equation}
C_c(v) = \frac{n - 1}{\sum_{u : u \neq v} d(u,v)}
\end{equation}

Instead of summation we can take maximum of distances from the node to all others, obtaining another version of distance based centrality, called {\em eccentricity}. This characteristic is formally defined by
\begin{equation}
e_c(v) = \max_{u \in V(G)} d(v,u).
\end{equation}

Another example of a centrality notion, different from those based on distance, is the {\em eigenvector centrality}. This centrality connects the importance of a node with the importance of all its neighbors by determining eigenvectors of the adjacency matrix, i.e. it can be determined implicitly 
by~\cite{Bonacich1987power}

\begin{equation}
C_e(v) = \frac{1}{\lambda} \sum_{u \in V(G)} a_{v,u} C_e(u),
\end{equation}

where $\lambda$ is a constant. The global version again just averages over all nodes in similar way to the clustering coefficient.

It may seem that most global characteristics are just an average of local ones. There are, however, other characteristics describing some global property of a network that do not average over local characteristics. One example is the assortative character of the network, which represents the tendency of nodes to connect to other nodes with a similar degree~\cite{Newman2003Mixing}. The {\em assortative coefficient} is defined as

\begin{equation}
r = \frac{\sum_{uv \in E} k_uk_v - \overline{k}}{\sum_{u,v \in E}\frac{1}{2}(k_u^2+k_v^2) - \overline{k}},
\end{equation}
where $\overline{k} = \frac{1}{m}\left[ \sum_{u,v \in E}\frac{1}{2}(k_u+k_v)\right]^2$.

\section{Nonlinearity assessment}\label{s:nonlinass}



In this section, we outline the proposed procedure for assessing the effect of nonlinearity on network properties. This procedure is demonstrated by applying it to the above described dataset of NYSE100. For purposes of this section we use $x_i$ to denote time series appropriately preprocessed as described in section~\ref{s:data}.

\subsection{Exploring bivariate and marginal non-Gaussianities}
\begin{figure}[ht!]
\begin{center}
 \includegraphics[width=0.8\linewidth]{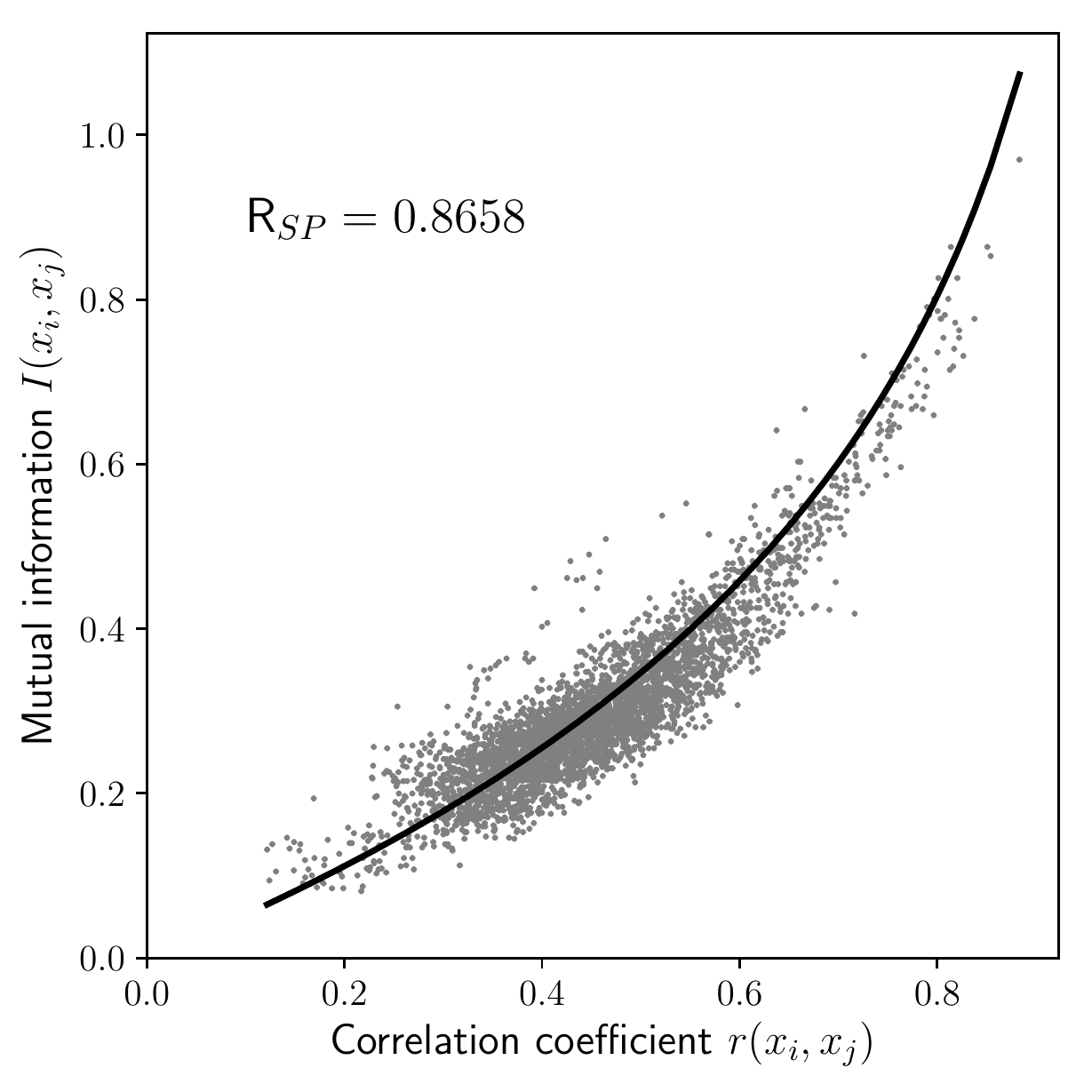}
\end{center}
\caption{Relation between correlation and mutual information estimates from the original data (log-returns of the close prices of the NYSE100 stocks) with the value of Spearman's correlation. Each gray point represents the values of $r(x_i, x_j)$ and $I(x_i, x_j)$ for a pair of variables $x_i, x_j;~i, j \in \{1,\ldots, N\}$. Gaussian mutual information $I_G$ shown by black line.}
\label{fig:r_vs_mi}
\end{figure}

The first step in exploration of nonlinearity of dependences is to visually inspect the deviation from linearity. A convenient tool for this purpose is the use of the scatter plot comparing the estimated values of both correlation and mutual information for each pair of variables. Inspection of Fig.~\ref{fig:r_vs_mi} shows that in our example dataset, mutual information generally grows with linear correlation, however this relation is not too tight. In particular, the mutual information estimate for many variable pairs deviates substantially from the theoretical value of the Gaussian mutual information $I_G$ (shown in black line).

\begin{figure}[ht!]
\begin{center}
 \includegraphics[width=0.8\linewidth]{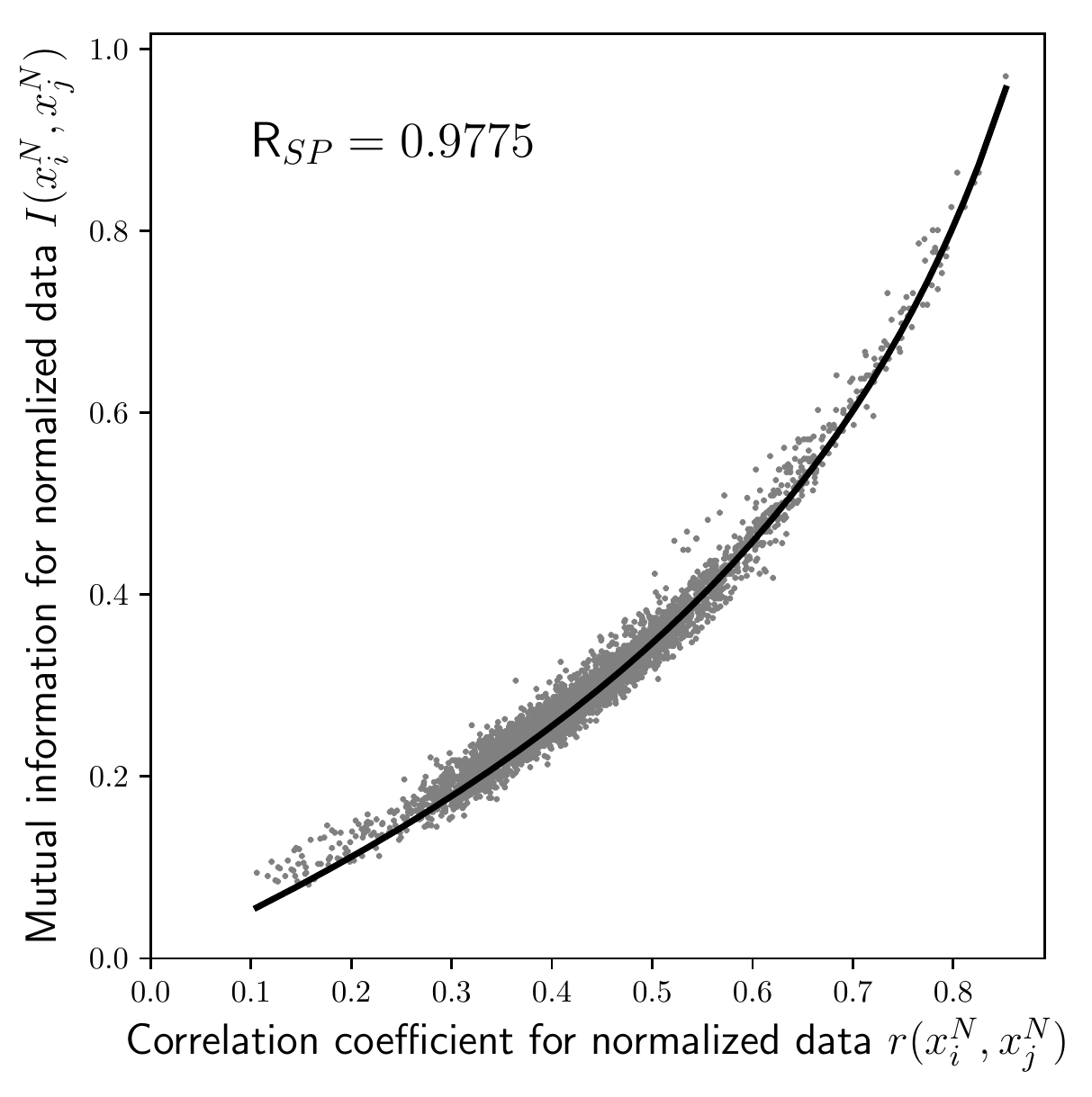}
\end{center}
\caption{Relation between correlation and mutual information estimates from the univariately normalized data. Visualization as in Figure~\ref{fig:r_vs_mi}.
}
\label{fig:normr_vs_normmi}
\end{figure}

As discussed earlier, the deviation from Gaussianity can be of two types; the first being given by non-Gaussianity of the univariate marginals and the second by non-Gaussianity of the bivariate dependence pattern (the copula). Notably, the former one can be removed by a simple monotonous rescaling that normalizes the marginal distributions. We denote time series preprocessed using this transformation as $x_i^N$. We also visualize the scatter plot after such rescaling in Fig.~\ref{fig:normr_vs_normmi}. After this correction, mutual information relates much more closely to correlation, suggesting that much of the mismatch was due to non-Gaussianity of the original univariate distributions (e.g. kurtosis, skewness, outliers,...).
From a practical viewpoint, this suggests that the evaluation of mutual information might be reasonably substituted by the computationally much less demanding operations of univariate normalization and evaluation of a linear correlation coefficient, or similarly by computation of Spearman's correlation coefficient from the original data.

\subsection{Quantyfing and localizing nonlinearity}
\begin{figure}[ht!]
\begin{center}
 \includegraphics[width=0.8\linewidth]{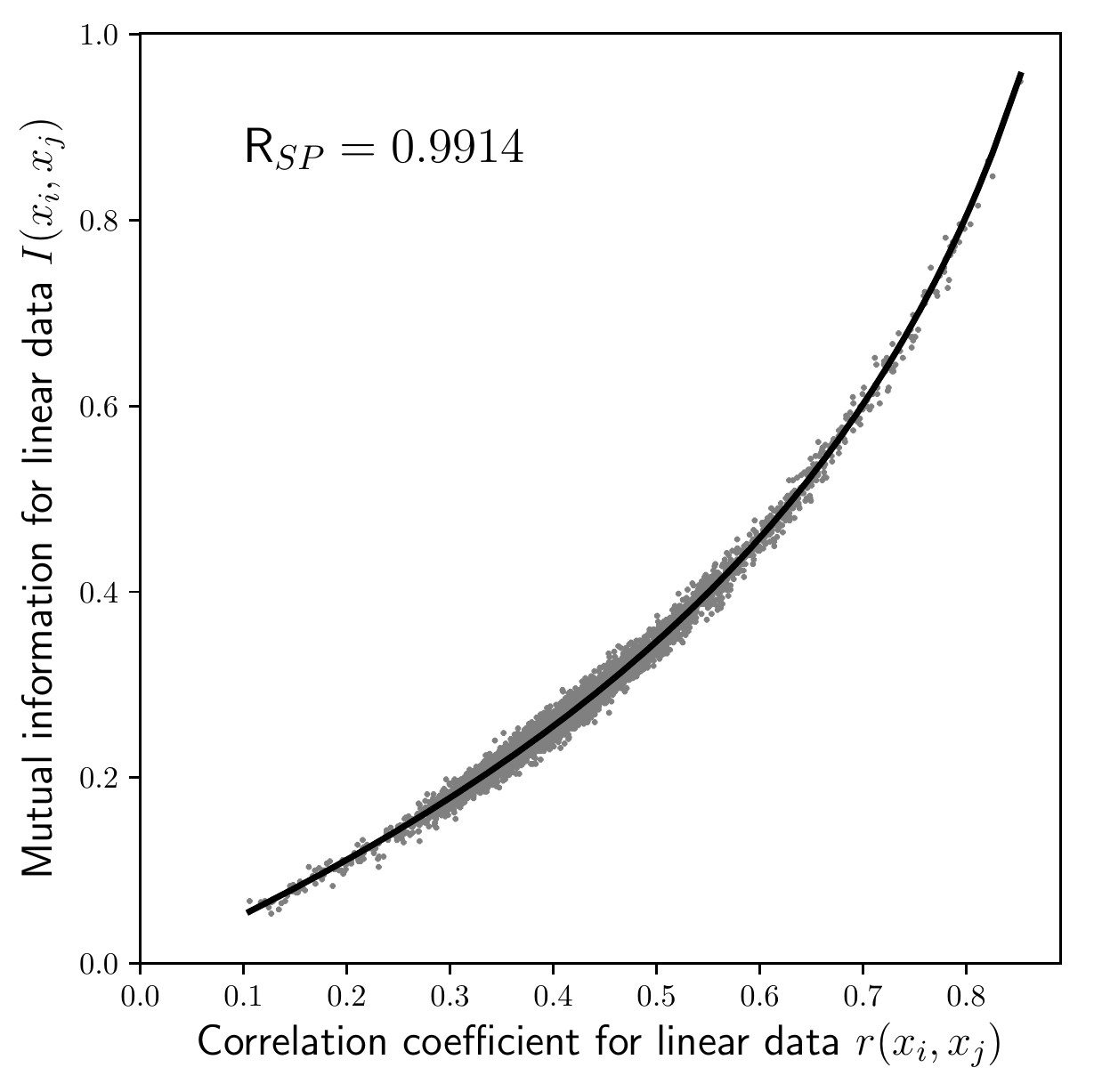}
\end{center}
\caption{Relation between correlation and mutual information estimates computed for linearized data constructed as surrogate dataset from original data. Visualization as in Figure~\ref{fig:r_vs_mi}.
}
\label{fig:surr_normr_vs_normmi}
\end{figure}

However, there is still some remaining mismatch between the correlation coefficients and mutual information in Fig.~\ref{fig:normr_vs_normmi}. We proceed by asking several questions: How substantial is this nonlinearity? What is the nature of this nonlinearity? And, finally: How does it affect the graph properties? We also note that some deviation between mutual information and correlation (or the corresponding Gaussian mutual information) may be due to inaccuracy of the estimates of mutual information and correlation from finite data samples, see also relation between correlation coefficient and mutual information computed for linearized data in Figure~\ref{fig:surr_normr_vs_normmi} -- making data linear has been performed using surrogate dataset, see description in section~\ref{ssec:NetProps}.

\begin{figure}[ht!]
\begin{center}
 \includegraphics[width=1.0\linewidth]{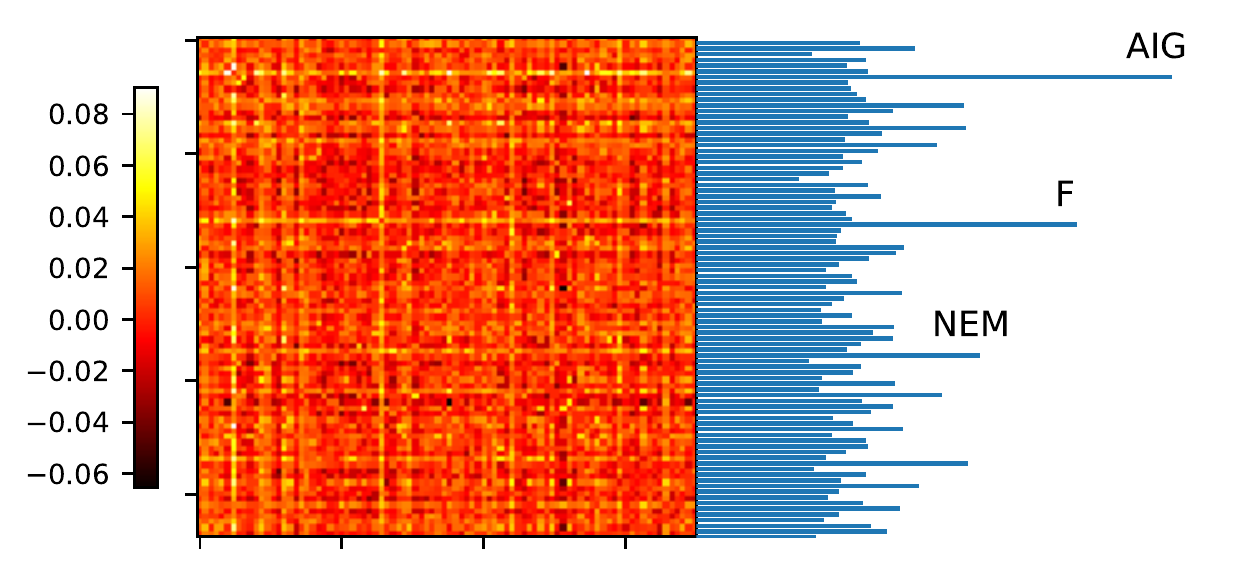}
\end{center}
\caption{Extranormal information estimates $I_e(x^N_i, y^N_i)$ for all pairs of variables $x^N_i, x^N_j;~i, j \in \{1,\ldots, N\}$ represented as $N$ times $N$ matrix with additional bar graph showing sums of columns representing overall nonlinearity in each stock.}
\label{fig:etranorm_mi_matrix}
\end{figure}

\begin{figure}[ht!]
\begin{center}
 \includegraphics[width=0.8\linewidth]{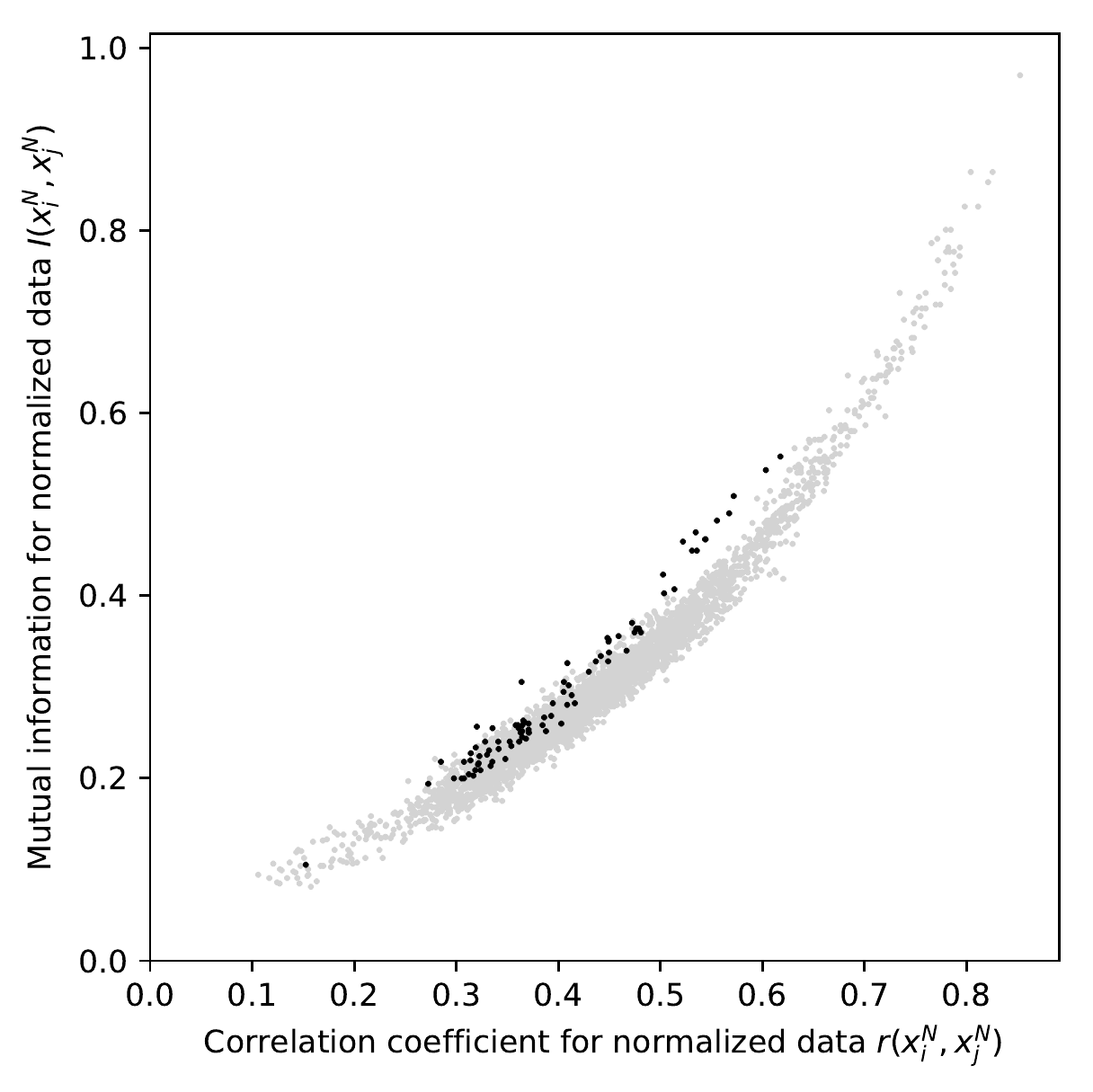}
\end{center}
\caption{Relation between correlation and mutual information estimates from the univari ately normalized original data. Selected data points that correspond to a stock with largest apparent nonlinearity (AIG) in black.
Visualization as in Figure~\ref{fig:r_vs_mi}.
}
\label{fig:normr_vs_normmi_enemies_selected}
\end{figure}

To quantify the observed nonlinearity, we compute an estimate of the extranormal information $\hat{I}_E$ for each pair of variables.
Fig.~\ref{fig:etranorm_mi_matrix} shows the extranormal information for all variable pairs for normalized data. One can clearly see that it is not homogeneously distributed, with some rows (columns) of the matrix containing markedly higher values than others. In this particular dataset we detected a marked outlier -- in the seventh row, representing the AIG stock. We can visualize the dependence values corresponding to this stock in the information/correlation scatter plot (see Fig.~\ref{fig:normr_vs_normmi_enemies_selected}).

\begin{figure}[ht!]
\begin{center}
 \includegraphics[width=1.0\linewidth]{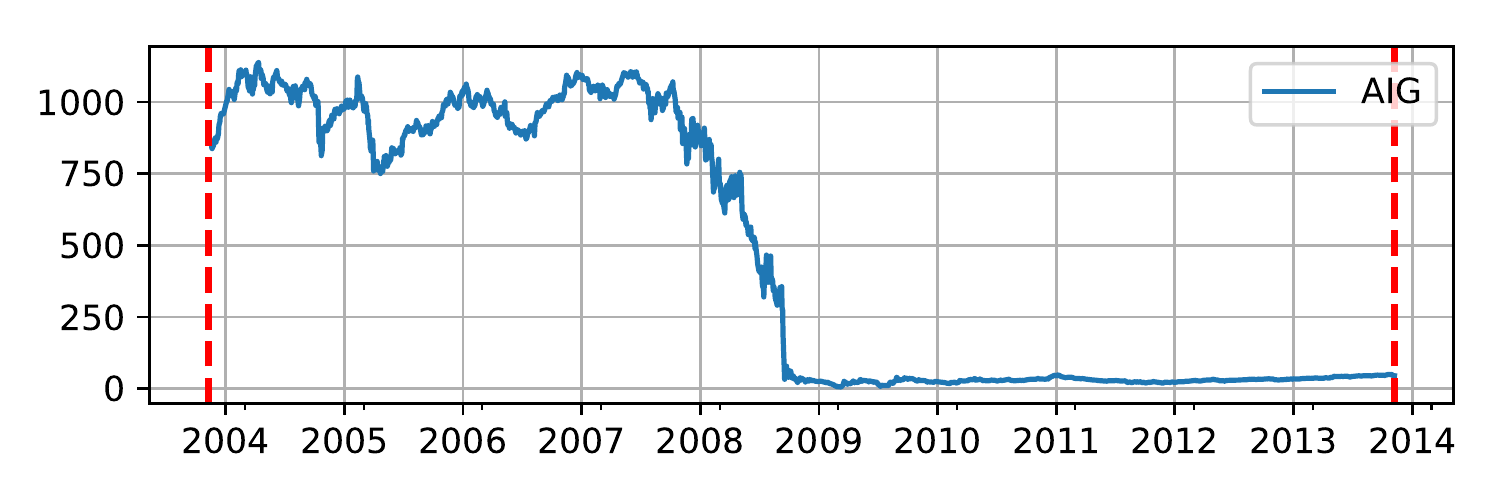}
\end{center}
\caption{Adjusted close price for a stock with the strongest apparent nonlinearity: the AIG stock.}
\label{fig:timeseries_AIG}
\end{figure}

One can see that indeed this stock is responsible for the most clear deviations between correlation and mutual information. Inspection of the corresponding time series (Fig.~\ref{fig:timeseries_AIG}) gives a nice example of a data feature that might lay behind what appears as a nonlinear dependence -- in particular a profound nonstationarity. As has been reported in a different context, apparent nonlinearity may be a useful sign of profound nonstationarities~\cite{Hlinka2013ClimDyn}. However, both correlation and mutual information estimation assumes (at least theoretically) stationarity, as such nonstationarities may lead to spurious detection of dependences. Based on this, we would suggest removal of such strongly nonstationary variables from the analysis, or at least interpreting the results concerning stationary estimates from heavily nonstationary data with care.

\subsection{Nonlinearity effect on network properties}
\label{ssec:NetProps}
The final step of the analysis is to assess whether the remaining nonlinearity has a substantial effect on the network properties. If this is the case, one may argue for the use of mutual information networks instead of correlation networks, as it contains meaningful added value. On the other hand, if the effect is not substantial, then correlation networks should be preferred, as the estimation of correlation is generally more simple, more robust (after necessary preprocessing steps mentioned above), and also easier to interpret and transfer the the findings.

In general, the effect of nonlinearity may depend on the type of graph construction method.
We start by studying the PMFG graph, and focus on the key network characteristics defined earlier: the closeness $C_c$, average clustering coefficient $C$, eccentricity $e_C$, characteristic shortest path $L$, assortative coefficient $r$, eigenvalue centrality $C_e$ and also average betweenness centrality $C_b$. The PMFG generated from the estimated mutual information matrix may differ from the PMFG of the correlation matrix for various reasons. One of them is a true nonlinearity of the underlying dependences, a second one is the random variation due to inaccuracy of the estimates of the mutual information from finite time series, and yet another one is the difference in the properties of the correlation and mutual information estimators which could have specific biases (e.g. the mutual information estimate might be more noisy, and therefore providing networks more similar to random networks). We avoid the latter problem by working with mutual information matrices of linear surrogate data instead of treating the correlation matrices directly, and evaluate whether the MI PMFG networks from data differ from those constructed from its linear surrogates. 

The linear surrogates conserve the linear structure (covariance and autocovariance) of the original data, but remove nonlinear structure from the multivariate distribution. Computationally, linear surrogate data can be constructed as multivariate Fourier transform (FT) surrogates \cite{Prichard1994generating,Palus1997detecting}. The procedure includes computing the Fourier transform of the series, keeping unchanged the magnitudes of the Fourier coefficients, adding a randomly selected number to the phases of coefficients within the same frequency bin; and finally applying the inverse FT into the time domain. Using this procedure, we construct the mutual information matrix and the corresponding graph for each of $N_S=999$ surrogate datasets constructed by this procedure. Then, the properties of the data-derived PMFG can be compared to those of the surrogate-derived PMFGs; any statistical deviation can be ascribed to a deviation of the data from linearity. Note that this step is not redundant with respect to the previous procedure of quantifying the nonlinearity for individual variable pairs, as one may conceive examples where weak couplings (or even only subtle bivariate nonlinearities) give rise to substantial effects at the network level and vice versa. 

\begin{figure}[ht!]
\begin{center}
 \includegraphics[width=1.0\linewidth]{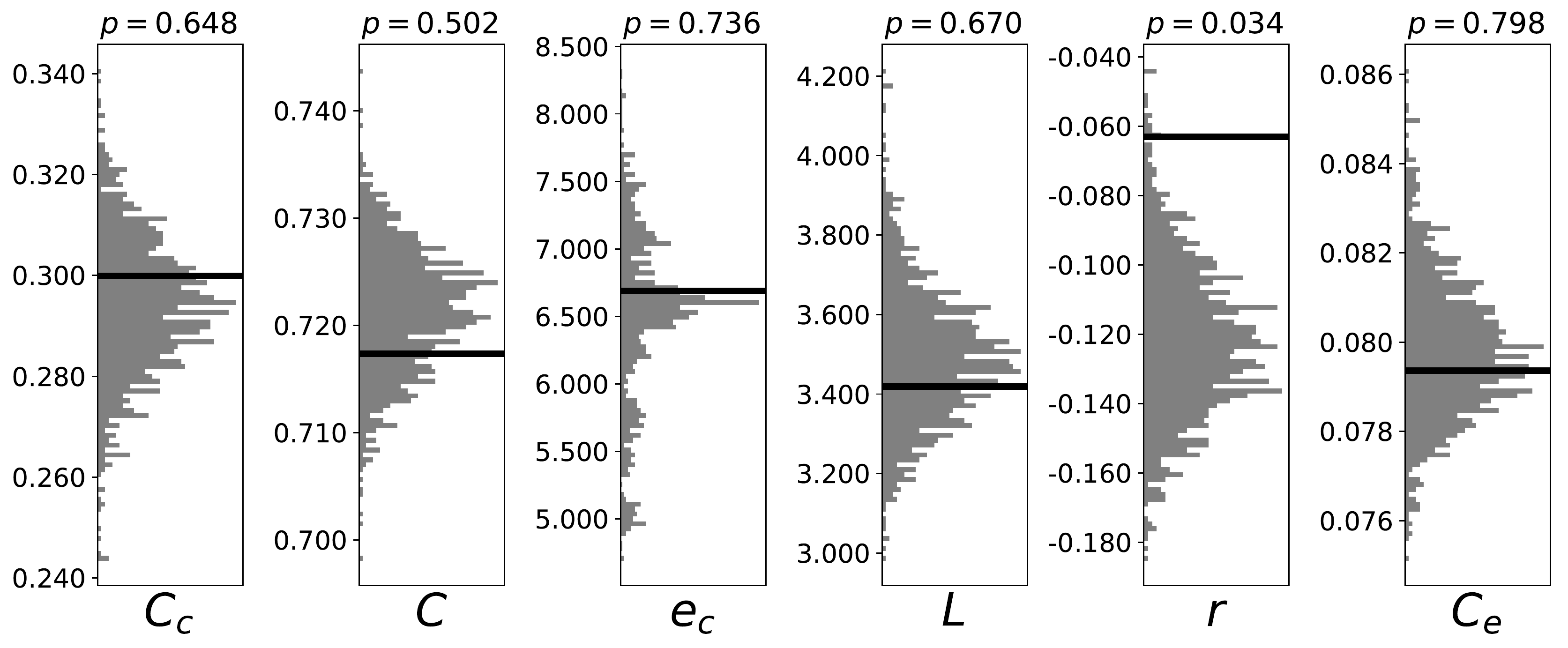}
\end{center}
\caption{Global graph characteristics of PMFG networks from original data with stock AIG removed (black line) and from corresponding linearized surrogate datasets (histogram in grey). Connectivity has been determined via mutual information and univariately normalized data were considered.  Presented characteristics are closeness $C_c$, average clustering coefficient $C$, eccentricity $e_C$, average shortest path $L$, assortative coefficient $r$ and eigenvalue centrality $C_e$. The presented p-values correspond to a two-sided non-parametric test against null distribution generated by linear surrogates.}
\label{fig:graph_char_norm_log_ret_noAIG1}
\end{figure}

The results of this analysis are shown in Fig.~\ref{fig:graph_char_norm_log_ret_noAIG1}. For all of the studied graph properties (with exception in the case of the assortativity $r$), the MI graph property does not deviate from the values obtained from linear data surrogates with same correlation matrix. Indeed, the MI graph values are not identical to the correlation graph values. However, the difference appears for all graph properties but the assortativity attributable to random variability of MI graph estimates. More rigorously, this comparison can be framed as statistical testing using an empirical null distribution (represented by the linear surrogate model realizations) of the graph properties under the hypothesis of multivariate normal distributions with given correlation structure. Given a comparison with 999 surrogates, we can evaluate the obtained significance level by the position of the data point in the observed surrogate distribution; in the absence of directional hypothesis, we report two-sided test results.

\begin{figure}[ht!]
\begin{center}
 \includegraphics[width=1.0\linewidth]{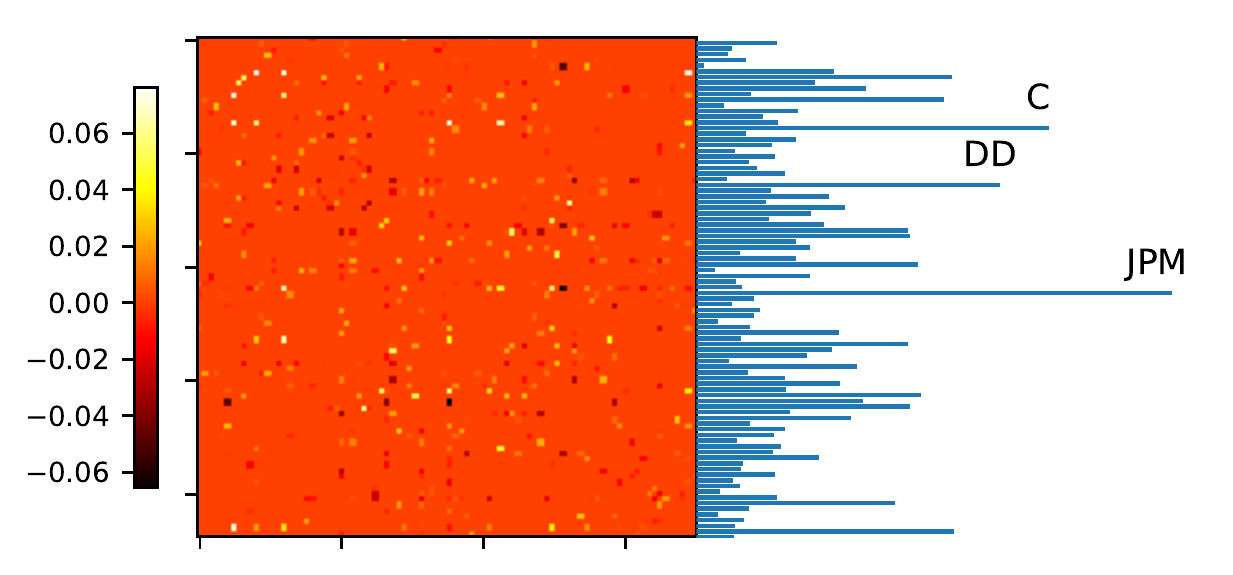}
\end{center}
\caption{Extranormal information estimates $I_e(x^N_i, y^N_i)$ for pairs of variables $x^N_i, x^N_j$ with $(i, j)$ an edge of the PMFG. Bar graph shows sums of columns representing overall nonlinearity in each stock within the PMFG.}
\label{fig:etranorm_mi_matrix_PMFG}
\end{figure}

\begin{figure}[ht!]
\begin{center}
 \includegraphics[width=1.0\linewidth]{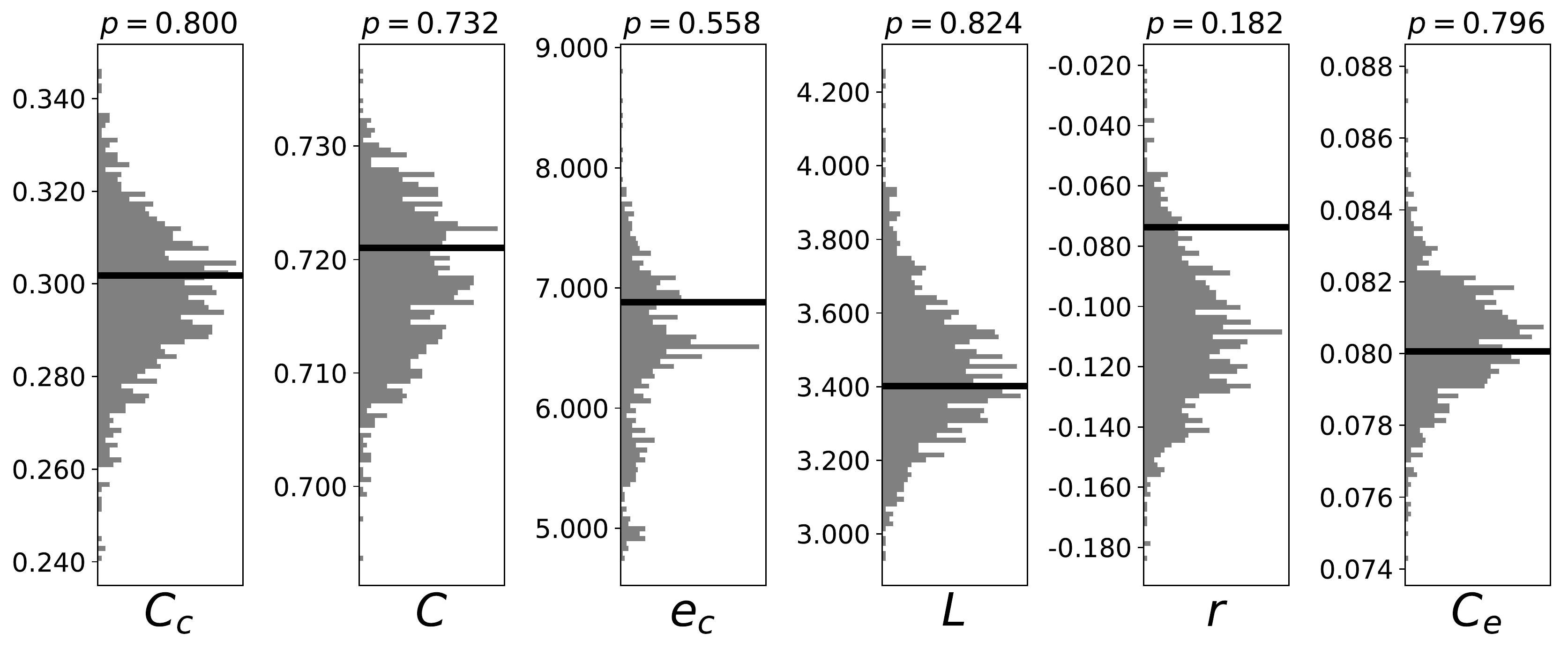}
\end{center}
\caption{Global graph characteristics of PMFG networks from original data with stocks AIG, C and JPM removed and from corresponding linearized surrogate datasets. Visualization and other settings as in Figure~\ref{fig:graph_char_norm_log_ret_noAIG1}.}
\label{fig:graph_char_norm_log_ret_noAIG1_JPM_C}
\end{figure}

We have observed a significant increase of the assortative coefficient $r$ in data compared to linear surrogates. 
To understand this property in more detail, we study the nonlinearities in more detail.
While AIG was marked as nonstationary based on sum of extranormal information over the whole matrix (see Figure~\ref{fig:etranorm_mi_matrix}), once we have decided to work with PMFG graph, the deviation from normality may be  more relevantly quantified by summing only over links in the correlation PMFG graph. Based on this analysis, shown in Figure~\ref{fig:etranorm_mi_matrix_PMFG}, relatively high apparent nonlinearity related to several further stocks is detected, including stocks C and JPM. The remaining network does not differ from linear surrogates, see Figure~\ref{fig:graph_char_norm_log_ret_noAIG1_JPM_C}. 

\begin{figure}[ht!]
\begin{center}
\includegraphics[width=1.0\linewidth]{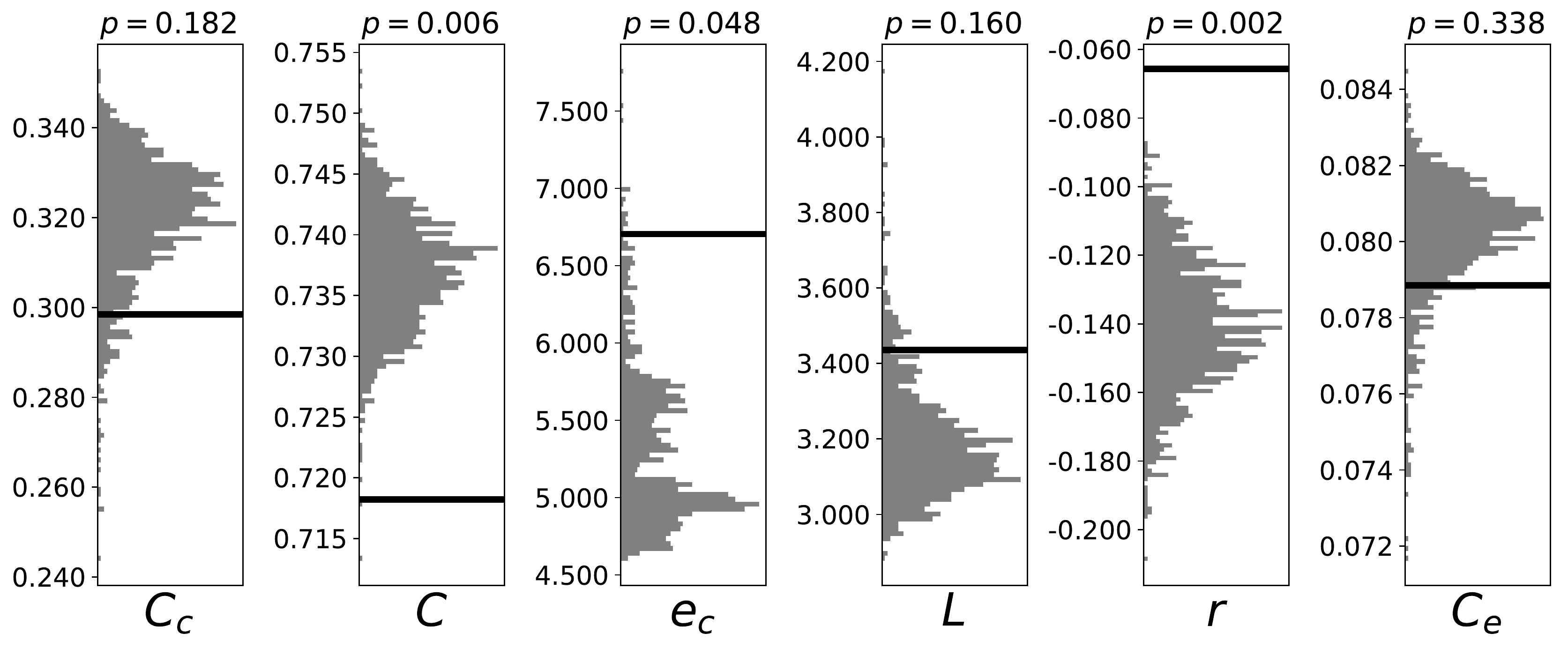}
\end{center}
\caption{Global graph characteristics of PMFG networks from original data and for linearized surrogate datasets.  Visualization and other settings as in Figure~\ref{fig:graph_char_norm_log_ret_noAIG1}.}
\label{fig:graph_char_log_ret}
\end{figure}

Notably, when the same analysis is applied to the original data before the univariate normalization and removal of nonstationary stocks, the picture is quite different, see Figure~\ref{fig:graph_char_log_ret}.
In particular, the MI graph has a clearly lower clustering coefficient compared to values observed for linear surrogate graphs (p=0.006). On the other side, there is higher eccentricity and assortative coefficient compared to linear surrogates. We can conclude that it is indeed the non-Gaussianity of the marginals (rather than non-Gaussianity of the dependence -- the copula) what is responsible for most of the mismatch between the linear correlation and the mutual information graph properties; some remaining differences are attributable to apparent nonlinearity due to nonstationarity.

\begin{figure}[ht!]
\begin{center}
 \includegraphics[width=1.0\linewidth]{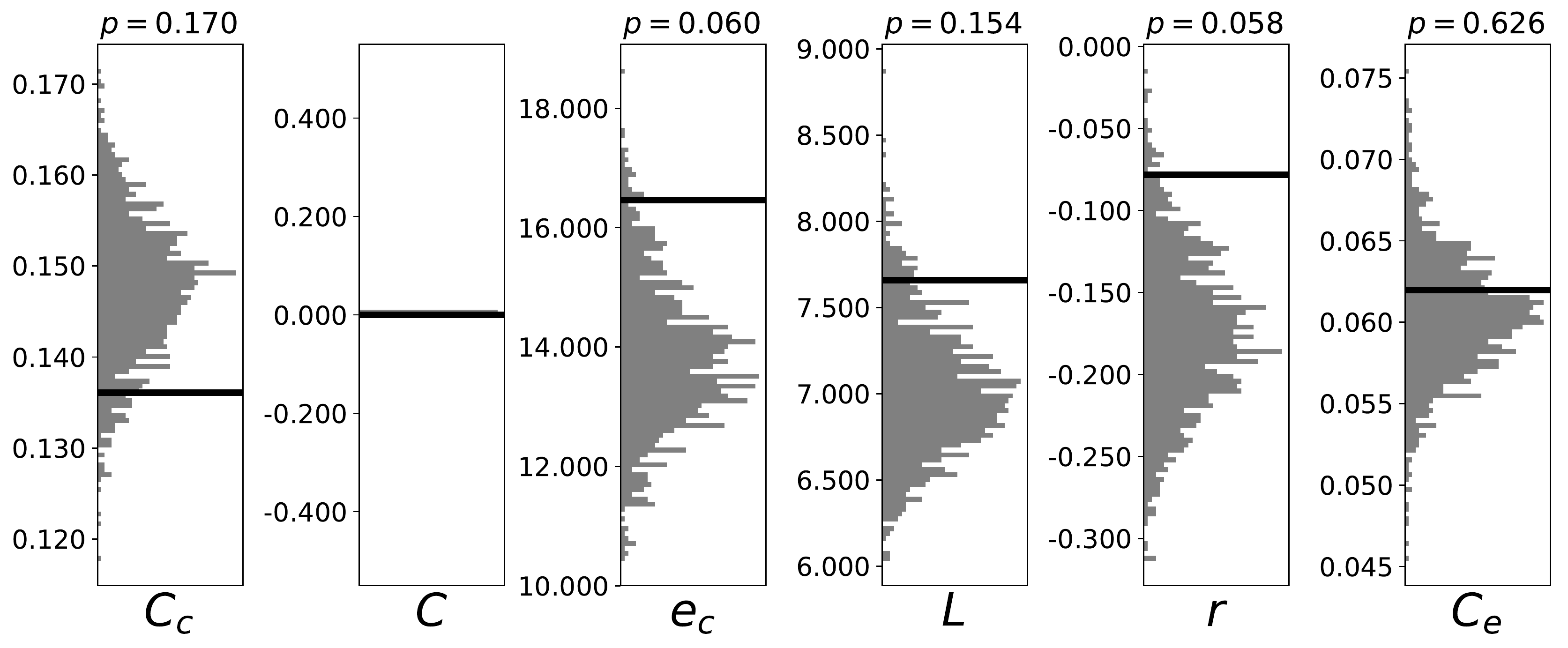}
\end{center}
\caption{Global graph characteristics of MST networks from original data with stocks AIG and BEN removed and from corresponding linearized surrogate datasets. Visualization and other settings as in Figure~\ref{fig:graph_char_norm_log_ret_noAIG1}.}
\label{fig:mst_graph_char_norm_log_ret_noAIG1_BEN_C}
\end{figure}

Another commonly used method for network construction is the minimal spanning tree (MST). Repeating the above described analysis, we do not observe significant deviation from the graph-theoretical structure corresponding to multivariate Gaussian data, with some tendency towards increased eccentricity and assortativity (see Figure~\ref{fig:mst_graph_char_norm_log_ret_noAIG1_BEN_C}).

\begin{figure}[ht!]
\begin{center}
 \includegraphics[width=1.0\linewidth]{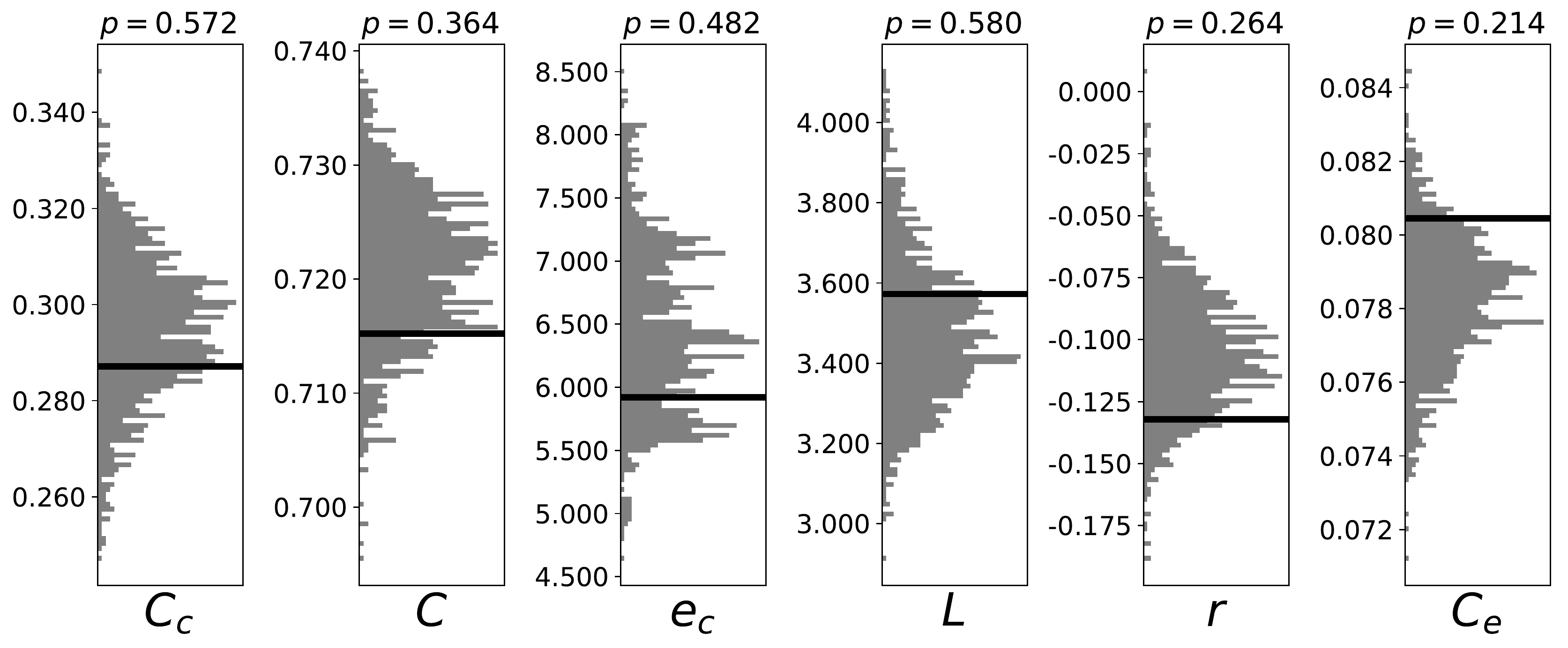}
\end{center}
\caption{Global graph characteristics of PMFG networks from original data and from corresponding linearized surrogate datasets for a shorter period from November 11, 2003 to May 11, 2008.  Visualization and other settings as in Figure~\ref{fig:graph_char_norm_log_ret_noAIG1}.}
\label{fig:graph_char_norm_log_ret}
\end{figure}

\begin{figure}[ht!]
\begin{center}
\includegraphics[width=1.0\linewidth]{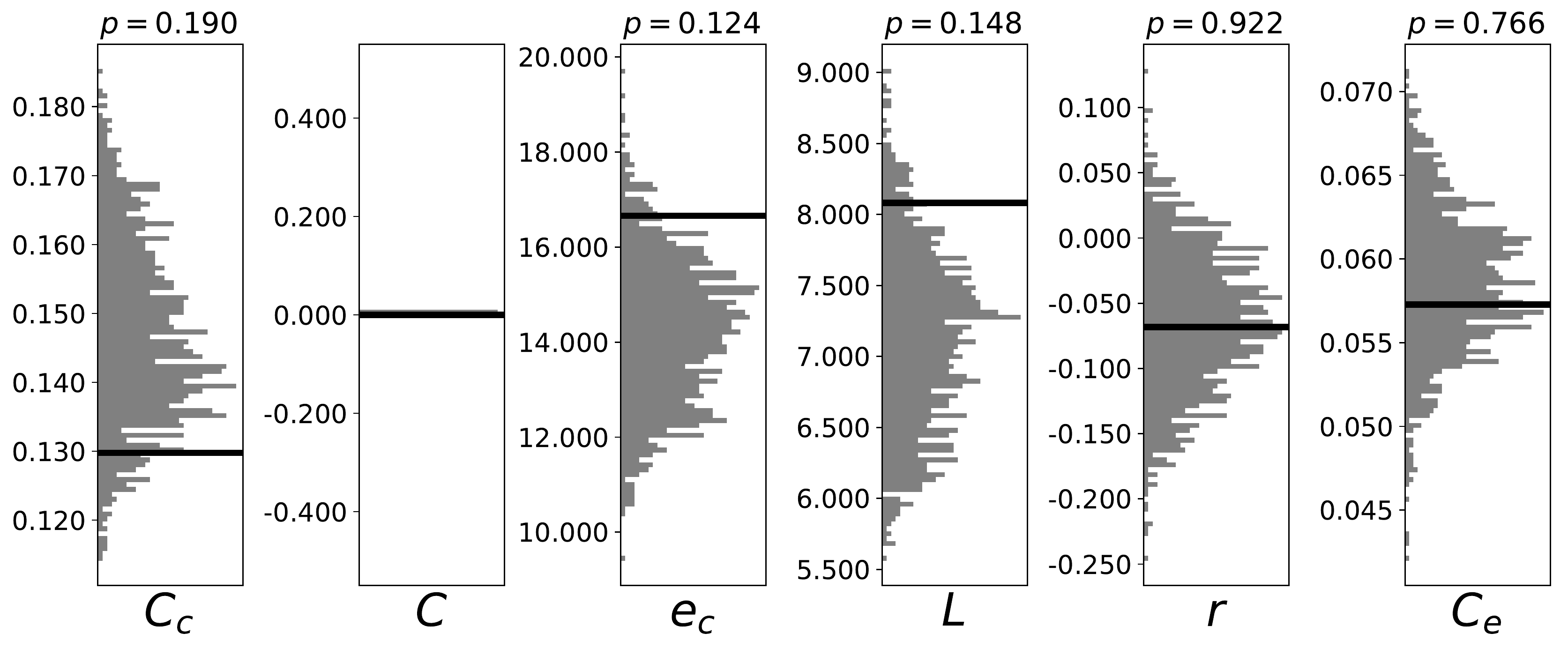}
\end{center}
\caption{Global graph characteristics of MST networks from original data and from corresponding linearized surrogate datasets for a shorter period from November 11, 2003 to May 11, 2008.  Visualization and other settings as in Figure~\ref{fig:graph_char_norm_log_ret_noAIG1}.}
\label{fig:mst_graph_char_norm_log_ret}
\end{figure}

Disregarding dependence estimates for nodes that strongly deviate from stationarity may be correct theoretically and also useful for stepwise localization of potential nonlinearities. However in practice, in cases when the whole network needs to be analyzed, it is more suitable to analyze only shorter segments of the data, as the nonstationarity may not be prominent on such shorter time scale. Inspection of our data suggest that most of the observed nonstationarity might be related to the global financial crisis in 2008. This is in line with the documented change of network pattern around this crisis~\cite{Heiberger2014}. When only data from a shorter period before the crisis is considered, the mutual information network properties do not substantially deviate from those of linear surrogate data, see Figure~\ref{fig:graph_char_norm_log_ret} and Figure~\ref{fig:mst_graph_char_norm_log_ret}.

\begin{figure}[ht!]
\begin{center}
 \includegraphics[width=1.0\linewidth]{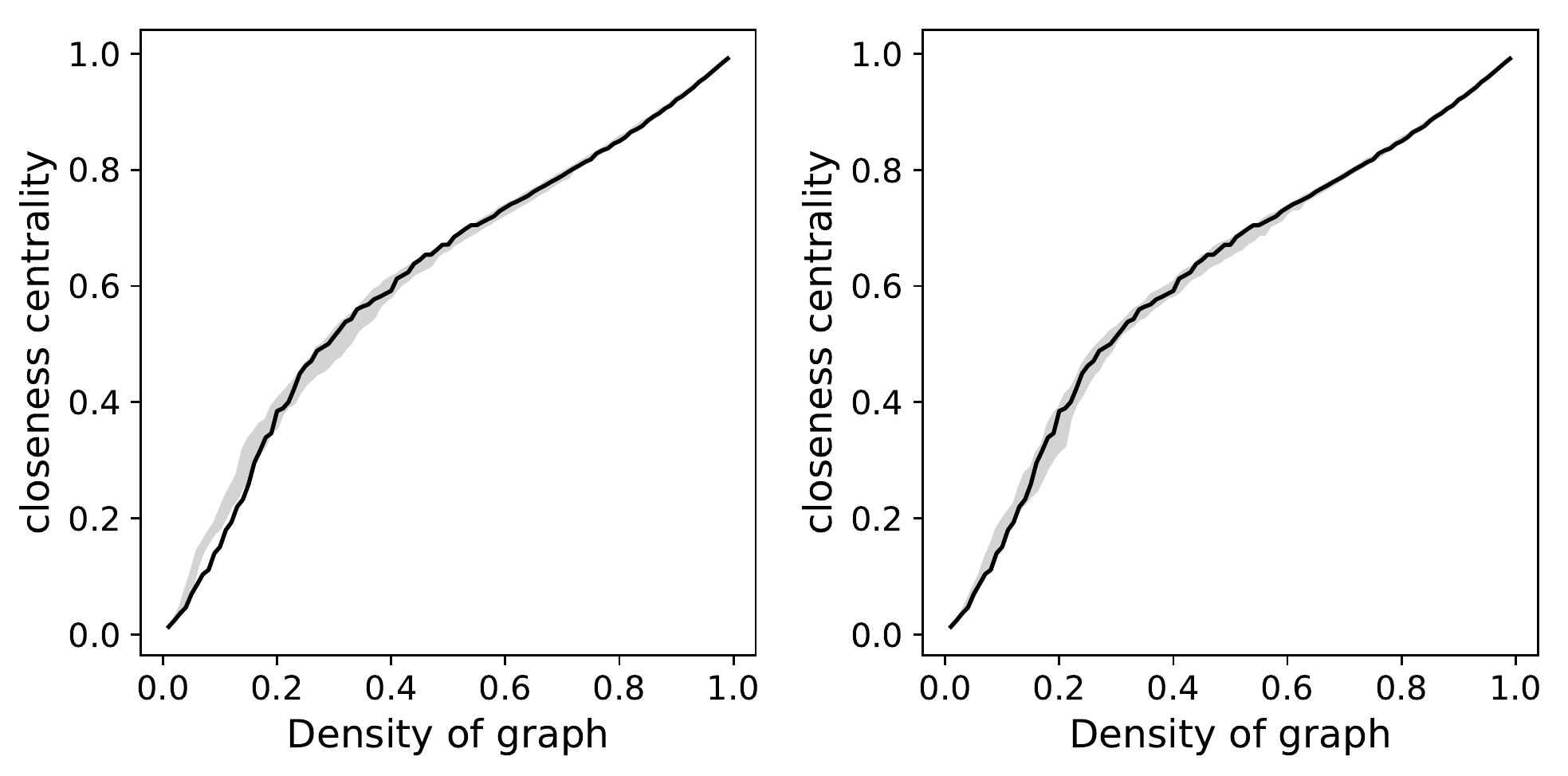}
\end{center}
\caption{Closeness centrality as a function of density for a network determined via winner-takes-all filtering. Plots are data derived networks (black lines) and gray area representing interval of plots of linearized surrogate datasets (gray area). The network is constructed from the real data with the stock AIG removed. Connectivity has been computed using mutual information without any normalization (left) and with marginal normalization applied (right).
}
\label{fig:clustering_wta}
\end{figure}

\begin{figure}[ht!]
\begin{center}
 \includegraphics[width=1.0\linewidth]{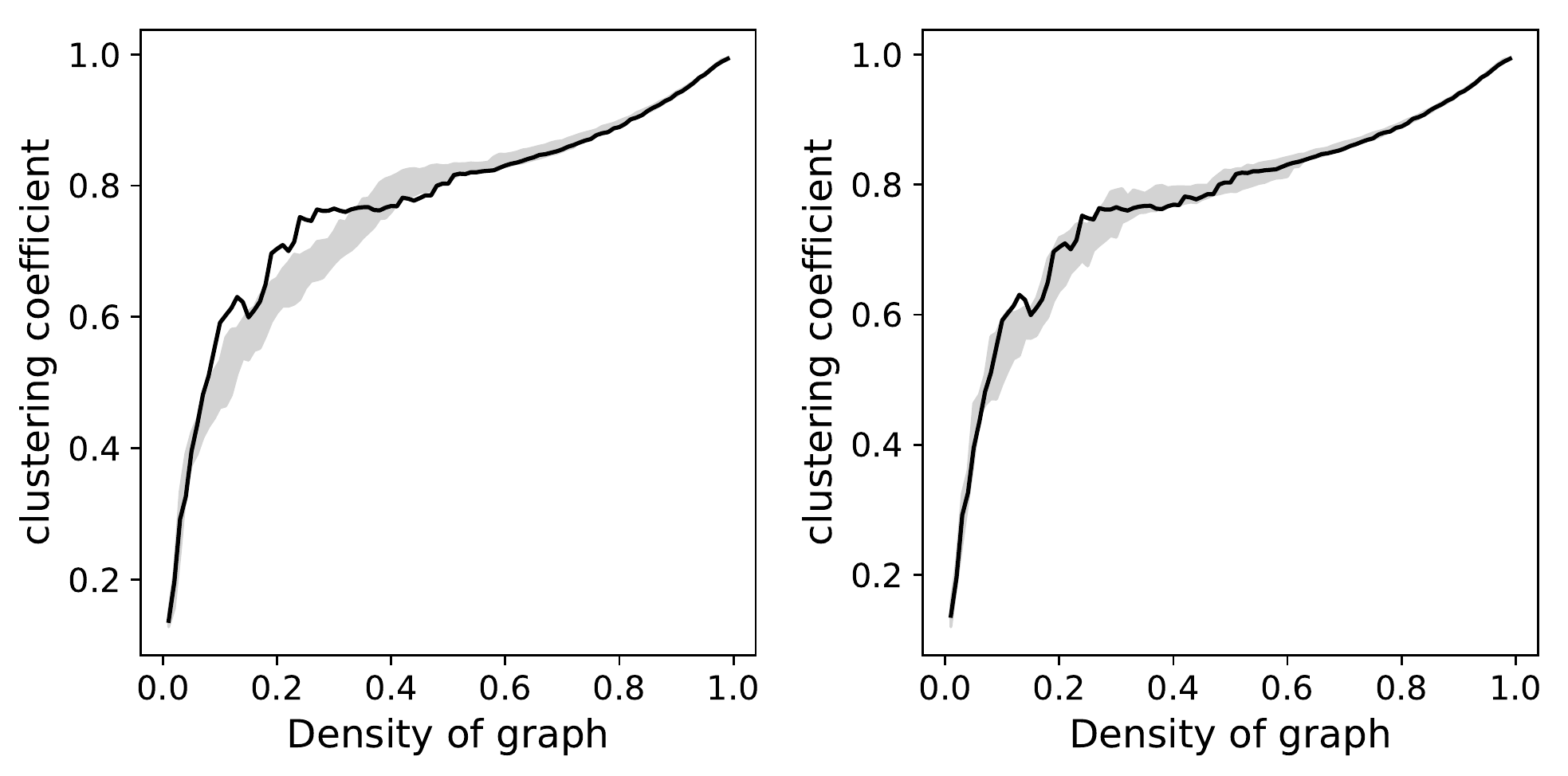}
\end{center}
\caption{Clustering coefficient as a function of density for WTA filtering; visualization as in Figure~\ref{fig:clustering_wta}.
}
\label{fig:c_wta}
\end{figure}

\begin{figure}[ht!]
\begin{center}
 \includegraphics[width=1.0\linewidth]{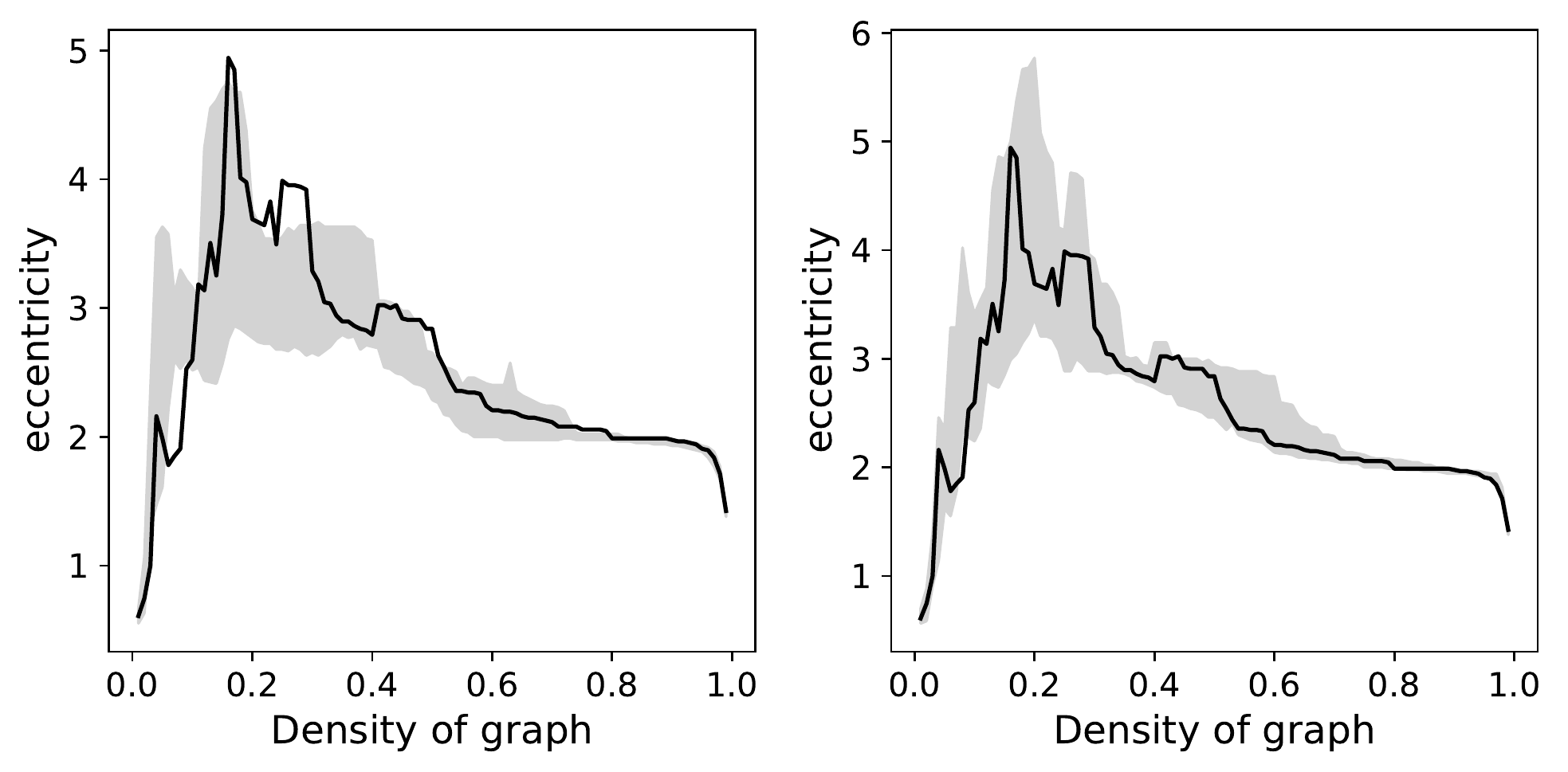}
\end{center}
\caption{Eccentricity as a function of density for WTA filtering; visualization as in Figure~\ref{fig:clustering_wta}.
}
\label{fig:ecc_wta}
\end{figure}

\begin{figure}[ht!]
\begin{center}
 \includegraphics[width=1.0\linewidth]{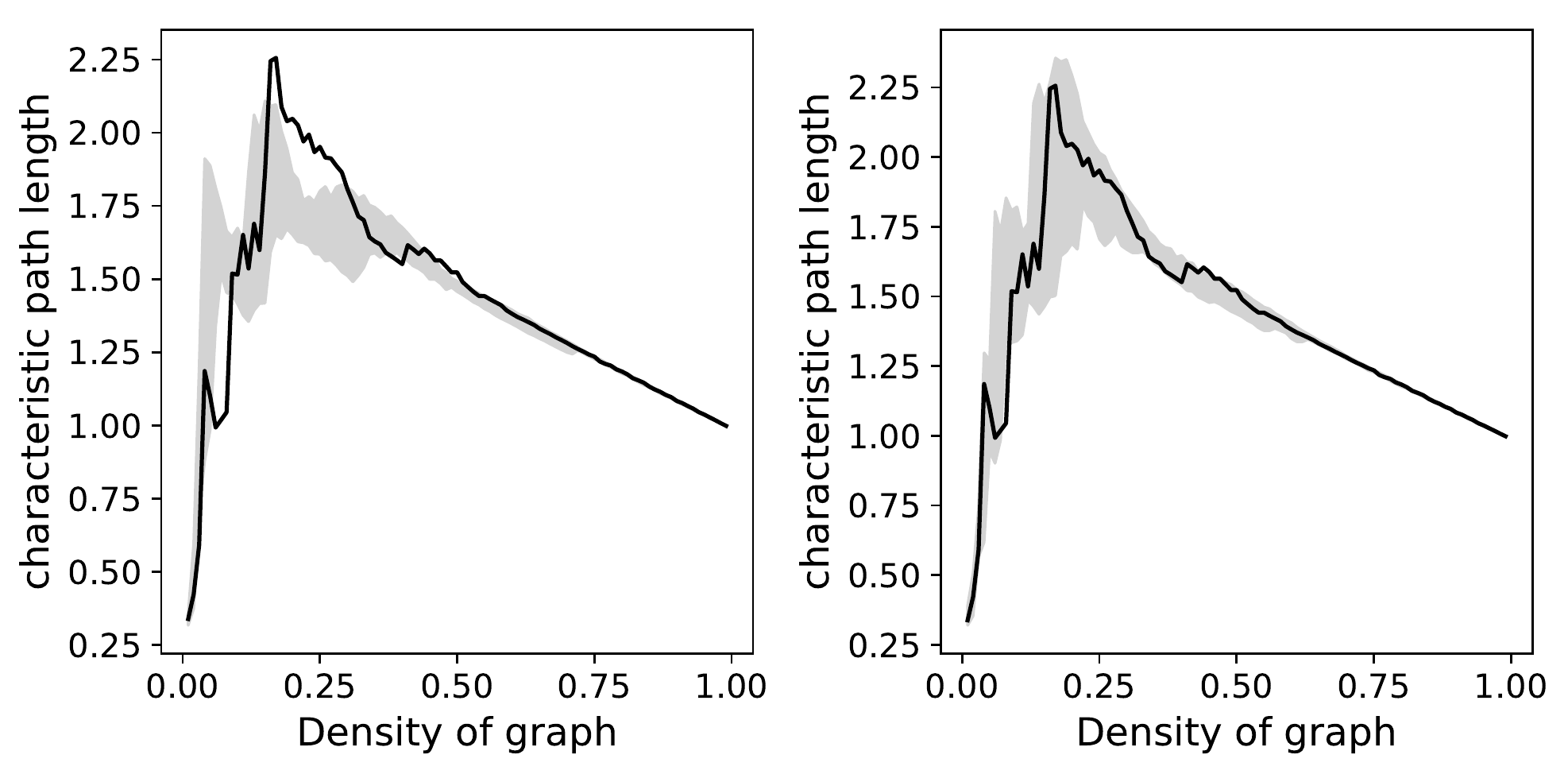}
\end{center}
\caption{Characteristic path length as a function of density for WTA filtering; visualization as in Figure~\ref{fig:clustering_wta}.
}
\label{fig:asp_wta}
\end{figure}

\begin{figure}[ht!]
\begin{center}
 \includegraphics[width=1.0\linewidth]{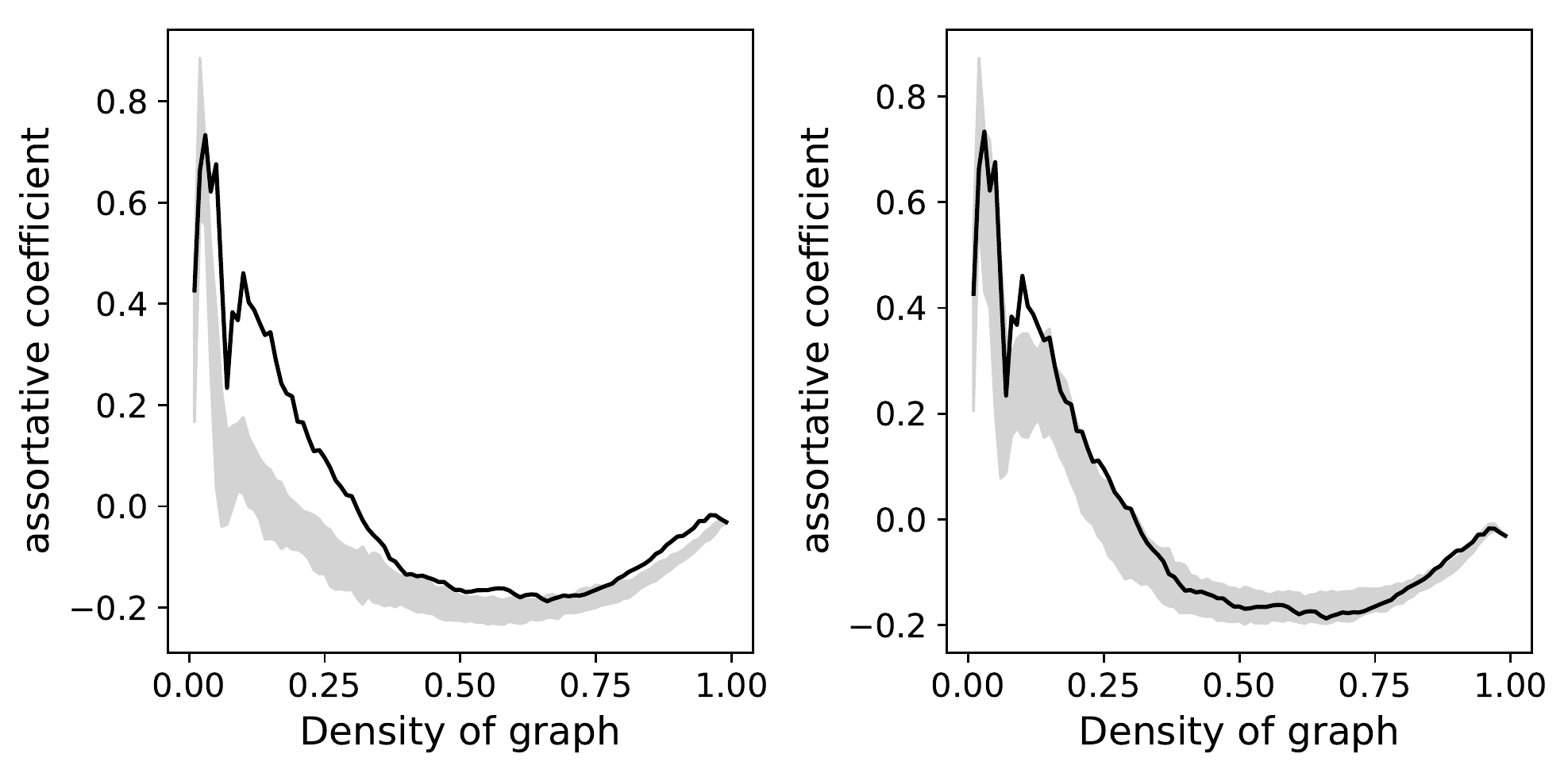}
\end{center}
\caption{Assortative coefficient as a function of density for WTA filtering; visualization as in Figure~\ref{fig:clustering_wta}.
}
\label{fig:r_wta}
\end{figure}

\begin{figure}[ht!]
\begin{center}
 \includegraphics[width=1.0\linewidth]{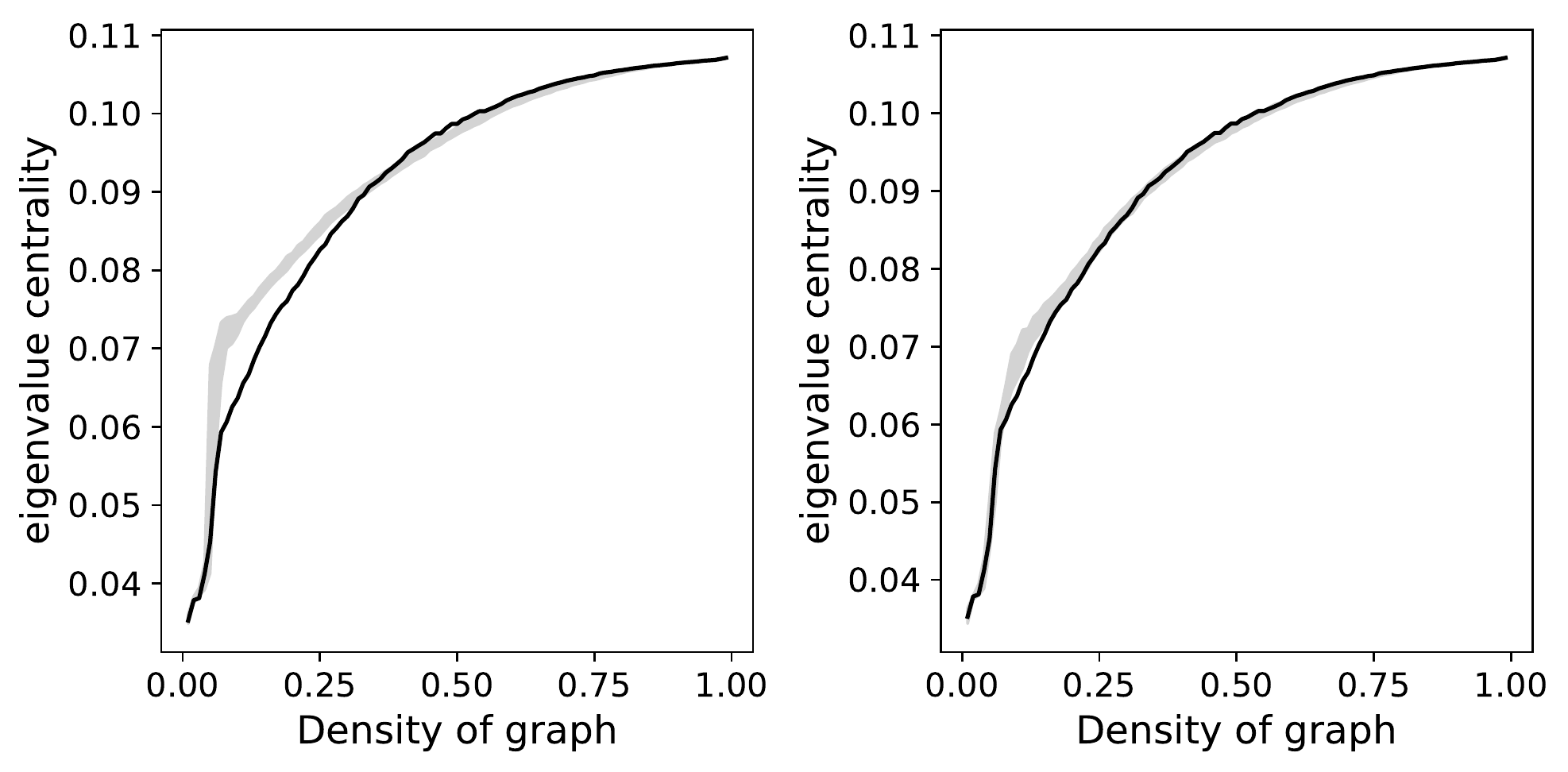}
\end{center}
\caption{Eigenvector centrality as a function of density for WTA filtering; visualization as in Figure~\ref{fig:clustering_wta}.
}
\label{fig:eig_wta}
\end{figure}

Of course, using the PMFG or MST are options for network construction, leading to a very sparse representation of the network. A commonly used alternative method of network construction providing principally arbitrary network density is to use the WTA approach, i.e. choosing a pre-defined threshold to obtain an unweighted graph using thresholding procedure that retains only some proportion of the strongest links, with the density of the obtained graph being a free parameter of the method. In theory, the effect of nonlinearity might be more important for some densities of the graph, so we carry out the analysis across a range of densities from 0 to 0.99 with a step of 0.01. The results are shown in Figures~\ref{fig:clustering_wta},~\ref{fig:c_wta},~\ref{fig:ecc_wta},~\ref{fig:asp_wta},~\ref{fig:r_wta}, and~\ref{fig:eig_wta}. Note that for all the graph properties (with some exemption for assortativity), the nonlinearity does not have a significant effect on the graph properties, across a full range of the network thresholds. Similarly as in the case of the PMFG graph, this is not so for the original data before marginal normalization, suggesting again the crucial importance of this preprocessing step.


\section{Discussion}\label{s:discussion}

We have outlined a pipeline to remove the effects of supposed nonlinearity and possibly localize any remaining components of nonlinearity in multivariate stock data when constructing the corresponding complex network. This pipeline has been demonstrated on the set of 100 U.S. stocks included in the NYSE100 index. We have obtained similar results by applying the procedures to sets of stocks from two other indices, namely the 100 stocks traded on the London Stock Exchange FTSE100 and 500 American stocks traded on the NYSE and NASQAD included in the index SP500; see Supplementary information.

The suggested procedure starts with the key step of marginal normalization. This was previously shown to be a very useful preprocessing step being able to remove most of the alleged nonlinearity in other datasets~\cite{Hlinka2011NeuroImage, Hlinka2013ClimDyn}, and this was confirmed also for stock networks. There are however some remaining contributions that could be explored using extranormal information. We provide an approach to localize these components and suggest their removal as they have shown to be mostly caused by nonstationarities, as shown for the example of the AIG stock. After the described preprocessing, along with the removal of outliers such as the AIG stock, most of the graph characteristics computer for PMFG and MST networks are systematically similar for networks constructed from linear and nonlinear data. Similarly, for series of parametrized networks created using WTA approach, the  described preprocessing removes differences of characteristics computed on WTA networks created from data and linear surrogates and thresholded for the same densities range. 
Theoretically, the presence of variables with outliers or nonstationarities should affect the WTA networks less then the PMFG or MST networks, as the presence of the links is mutually independent in the WTA approach (particularly, if a fixed threshold rather than fixed density is used). Therefore, the WTA networks should be more robust with respect to presence of such variables in the analysis.
Apart from fixed density of threshold, also a predefined significance threshold can be applied in the WTA networks. Note that for time series with strong autocorrelation, there is a bias towards higher amplitude of correlations, and an alternative thresholding procedure taking this effect into account was recently proposed~\cite{Palus2011discerding}. 


It is worth mentioning that after reducing the alleged nonlinearity behavior using the described pipeline there still remains some nonlinearity.
While testing nonlinearity of individual links can be also carried out, see e.g. Refs.~\cite{Hlinka2011NeuroImage,Hlinka2013ClimDyn}, statistical testing of each individual link generally leads to a large amount of tests (and related problems with multiple testing), and nevertheless the statistical significance of the individual nonlinearities may not be a good proxy of their amplitude or network relevance. For example, significant nonlinearity may be detected even for weak links with no effect on the global network structure. Thus, we have not reported statistical tests of linearity for each of the bivariate dependendes, but rather provide statistical inference related to the effect of nonlinearity on the key graph-theoretical properties of the network.

At this level, these residual contributions generally appear not to have significant effects on graph theoretical properties of constructed unweighted networks for most common thresholding strategies. As a practical result we can formulate the following proposition. When constructing unweighted stock networks, use correlation on marginally normalized data, or Spearman's rank correlation, which is almost equivalent. It is reasonable to be careful concerning nonstationarities and consider removing outliers or running on shorter segments of more stationary data. In summary, while there exists no universal solution for all datasets and the appropriate decision whether or not to use nonlinear measure always depends on particular analytic question and used dataset, for common stock network the utilization of linear correlation can be suggested. When dealing with other datasets than those presented in the current paper, when in doubt concerning suitability of application of nonlinear methods, we suggest to test always for presence of nonlinearity at least at the level of pairwise dependences, as graph-theoretical properties are often computationally demanding and have unstable estimates. 

In the current study, we have used a box-counting algorithm based on marginal equiquantization method~\cite{Palus1993information}. This algorithm is parametrized by number of bins desribing the level of discretization of the originally continuous variables. We have used number $4$ as suggested in literature~\cite{Fiedor2014networks,Fiedor2014Frequency}. As the recommended upper bound for bin count for this sample size is $13.765$, and there are some previous results using also $8$ bins in similar settings~\cite{NavetChen2008predictability}, we have also included results for $8$ bins in the Supplementary material. These showed similar properties to results obtained when using $4$ bins. Of course, there exists a range of other methods for mutual information estimation, which have not been used so far extensively in stock network literature, such as the $k$-nearest neighbor~\cite{Kraskov2004Estimating} or kernel methods~\cite{Moon1995estimation}.

To provide some evidence concerning the method's ability to quantify the nonlinearity with the current settings (sample size $N=2608$,  marginal equiquantization method with $4$ bins), we apply it to simulated data with a toy nonlinear distribution. In particular, we consider bivariate copula, where the sample $x,y$ is drawn with probability $p=0.5$ uniformly from $(0,0.5)\times(0,0.5)$, and with probability $1-p=0.5$ uniformly from $(0.5,1)\times(0.5,1)$. Here, the mutual information (in bits) is $I(X,Y)=1$ and the Gaussian information $I_G=\simeq 0.375$, leading to the non-Gaussian information $I_E\eqsim 0.625$. Simulating an ensemble of $1000$ realizations, we obtained estimates of the non-Gaussian information with mean $0.687$ and standard deviation $0.072$, i.e. clearly detecting the nonlinearity, while slightly overestimating its strength. More detailed analysis showed, that this is mainly due to the mutual information estimator being sligtly biased -- a known issue in estimating mutual information across methods, given by the imprecise approximation of the underlying distribution (in this case by the binning procedure).

To show robustness of the procedure and results we have shown its applicability to various datasets. For a fixed size, namely 100 stocks, we have tested two possible datasets given by stocks from indices FTSE100 and NYSE100. Results look similar in both cases. Moreover, these stocks are traded at different stock exchanges ensuring thus more reliable results of the method. To assess also larger size of dataset we included stocks from index SP500. Results also show similar patterns, see Supplementary material for detailed results.

The presented results constitute an extension of the previous studies into the role of nonlinearity in stock networks~\cite{Fiedor2014networks,kaya2015eccentricity,Baitinger2017interconnectedness2}. While these have described the theoretical motivation for the application of mutual information, and gave examples of both similarity and differences in the dependence structure and networks properties, we go further by systematically evaluating the nature and relevance of these observed differences. In particular, we suggest to consider separately non-Gaussianity of the copula and of the marginals (which can be treated by standard simple approaches without invoking information theory, such as rescaling marginals to normal distribution, or using rank correlation), provide a tool for quantification of the nonlinearity of dependences, and propose procedures to localize and diagnose the sources of non-Gaussianity in the data. The application to three well known stock indices has shown that the non-Gaussianity present is relatively weak and mostly attributable to marginal distribution corrigible by rescaling or to nonstationarities in the time series, particularly when spread across the 2008 global finance crisis. When these methodological points are taken into account, we have not observed substantial deviation from linear networks in the overall structure and across a range of network properties. 

Of course, this is not to mean that the observed processes are linear; we only suggest that a careful analysis of the observed data allows to discern and potentially mitigate several sources of apparent nonlinearity in stock index data, providing another step deeper into understanding of the system's structure, and allowing, at least partially, to rehabilitate the old and simple, but very useful and robust tool for stock network construction: the Pearson's linear correlation.

Finally, the presented procedures not only shed more light on the effect of nonlinearity in the stock networks at hand, but also are in principle applicable to other complex networks constructed for multivariate system characterized via measured collection of time series~\cite{Gao2016complex}.

\section{Supplementary material}
See Supplementary material for the results of further analysis described in the Discussion section.

\begin{acknowledgments}
We thank Milan Palu\v{s} for useful comments regarding presentation of the current study.
\end{acknowledgments}

\bibliography{references}

\end{document}